# Knowledization: Claim-Level Epistemic Control with Admissibility, Source Support, and Real-World Proxy Boundaries

Yan Jiao[1], Pin-Han Ho[1*], Limei Peng[1,2*], Yiming Miao[1]

[1]Shenzhen Institute for Advanced Study, University of Electronic Science and Technology of China.
[2]Kyungpook National University.

*Corresponding author(s). E-mail(s): pinhanho71@gmail.com; auroraplm@gmail.com;

**Abstract**

A plausible prediction need not be structurally identifiable, correctly grounded, supported by its attributed source, or warranted for promotion. We introduce KNowLEDIzATIoN, a claim-level epistemic-conformance and control architecture that keeps structural admissibility, realized support, source attribution, computational lineage, certification, and promotion explicit and non-substitutable. Quotient factorization supplies the exact global criterion, assumption-qualified Fisher geometry supplies a local diagnostic, and linear estimability provides an exact special bridge. Controlled interventions support the predicted progression from acquisition-dependent distinctions to claim-selective admissibility, source-separated belief updating, and finite ***K/I*** registry dynamics. Operational studies reveal a diferent boundary: scene-macro RH@32 reached **0.664** while exact grounding accuracy was **0.173**, and improved referential grouping reached **0.888** without meeting its prespecified **0.90** lower-confidence-bound criterion. High-fidelity simulation and natural-data studies therefore delimit operational realization rather than establish positive deployment efficacy. The contribution is a definition-complete composition and evaluation of established structural, evidential, provenance, and selective-control components.

# 1 Introduction

Modern prediction systems routinely attach scores to claims, but a score alone does not answer three prior questions: whether the observation process can distinguish the states on which the claim differs, which source supports the claim, and whether the available evidence warrants promotion beyond a tentative inference. Confidence and uncertainty estimation address important statistical questions [1–4], while selective prediction controls when a predictor abstains [5, 6]. Our focus is complementary: the epistemic contract connecting a claim to observation structure and declared evidence sources.

This problem intersects mature but differently scoped literatures. Identifiability and comparison of statistical experiments characterize recoverable distinctions; belief functions and multimodal evidence fusion model uncertainty, reliability, and conflict; database provenance records lineage; and reject-option, selective, and deferred prediction control abstention [7–14]. These components answer different questions. Our novelty claim is their disciplined composition at the claim level, not their individual invention:

$$\text{identifiable} \neq \text{supported} \neq \text{attributed} \neq \text{certified},$$
$$\text{source available} \neq \text{source supports this claim}.$$

Consider a distance claim produced from multimodal sensing. Removing a range channel can merge observation laws and destroy an exact distinction. Increasing finite Gaussian noise with the same structural sensing map can instead weaken finite-sample discrimination without changing which states have identical laws. A stored map or retrieved document may strongly favor the claim while contributing no observation-channel support. In all three cases a model can remain confident. Treating confidence, source availability, and claim-specific support as interchangeable obscures the reason a claim should be accepted or withheld.

The distinction is easiest to see by separating potential from realization. The declared acquisition and encoding process determines which distinctions could in principle be established. A particular observation then supplies realized support, and auxiliary sources may supply additional but differently attributed support. Thus an admissible claim can be weakly supported in one record, while an inadmissible claim can appear plausible because of prior or auxiliary information. Neither case is represented faithfully by a single confidence score.

A broader epistemic-integrity perspective places this separation within measurement theory. Task correctness and epistemic warrant are distinct: a correct claim can lack structural admissibility, realized support, valid source attribution, or traceable lineage, while an incorrect claim may still be well formed and auditable. A scalar confidence score cannot represent claim states that require different certification, attribution, or promotion decisions. Claim-level state, system-level competence, decision policy, and measurement instrument are therefore distinct objects; changing one does not establish a change in the others. Within that broader perspective, KNowLEDIzATIoN addresses criterion-referenced conformance of individual claims to a declared observation, evidence, and source contract. It does not construct a system-level competence measure. The unit of analysis is the claim record, not the system as a whole.

We study the end-to-end dependency sequence

$$
\begin{aligned}
\text{evidence/acquisition} &\longrightarrow \text{claim proposal/grounding} \\
&\longrightarrow \text{structural admissibility} \longrightarrow \text{source-aware support/attribution} \\
&\longrightarrow \text{certification} \longrightarrow \text{finite registries.}
\end{aligned}
$$

KNowLEDIzATIoN instantiates this sequence as an epistemic control layer rather than a replacement for a predictor. Its strongest exact object is the observation quotient: a claim is admissible precisely when it is constant on every observation-equivalence fibre. Local Fisher geometry provides a useful screen only when its null space is known to equal the fibre tangent. Support is recorded separately for the declared observation channel and auxiliary information. Computational lineage records origin and transformations; epistemic attribution separately records which source relation is credited with warrant. Certification has three states—certified, rejected, and unresolved—so lack of a sound decision procedure is not silently converted into rejection.

### *Semantic non-substitution.*

The architecture is governed by a design contract, not a new universal theorem: absent additional justified assumptions, confidence cannot substitute for support; lineage cannot substitute for epistemic attribution; source availability cannot substitute for realized claim support; local Fisher passage cannot substitute for global admissibility; support cannot substitute for structural admissibility; and support alone cannot authorize promotion. Certification and promotion are operational statuses, not philosophical certainty. The manuscript therefore uses "semantically separated" or "non-substitutable" rather than invoking a formal programming-language type system.

Figure 1 summarizes the two-branch architecture and keeps acquisition, realized support, lineage, attribution, and promotion in distinct semantic roles.

The paper addresses four research questions. The first three preserve the legacy experimental questions; the fourth collects the later operational and boundary studies.

- *RQ1* How do acquisition and encoding determine the global distinctions and local information structure available to an AI system?
- *RQ2* When are the distinctions required by a claim contained in those made identifiable by the declared observation experiment?
- *RQ3* How should structural admissibility, observation support, auxiliary support, and provenance jointly determine observation-certified knowledge, supported inference, and finite $K/I$ routing?
- *RQ4* How do finite claim proposal, grounding, learned support/attribution, and plausible real-world evidence proxies constrain an operational instantiation?

Our contributions are fivefold.

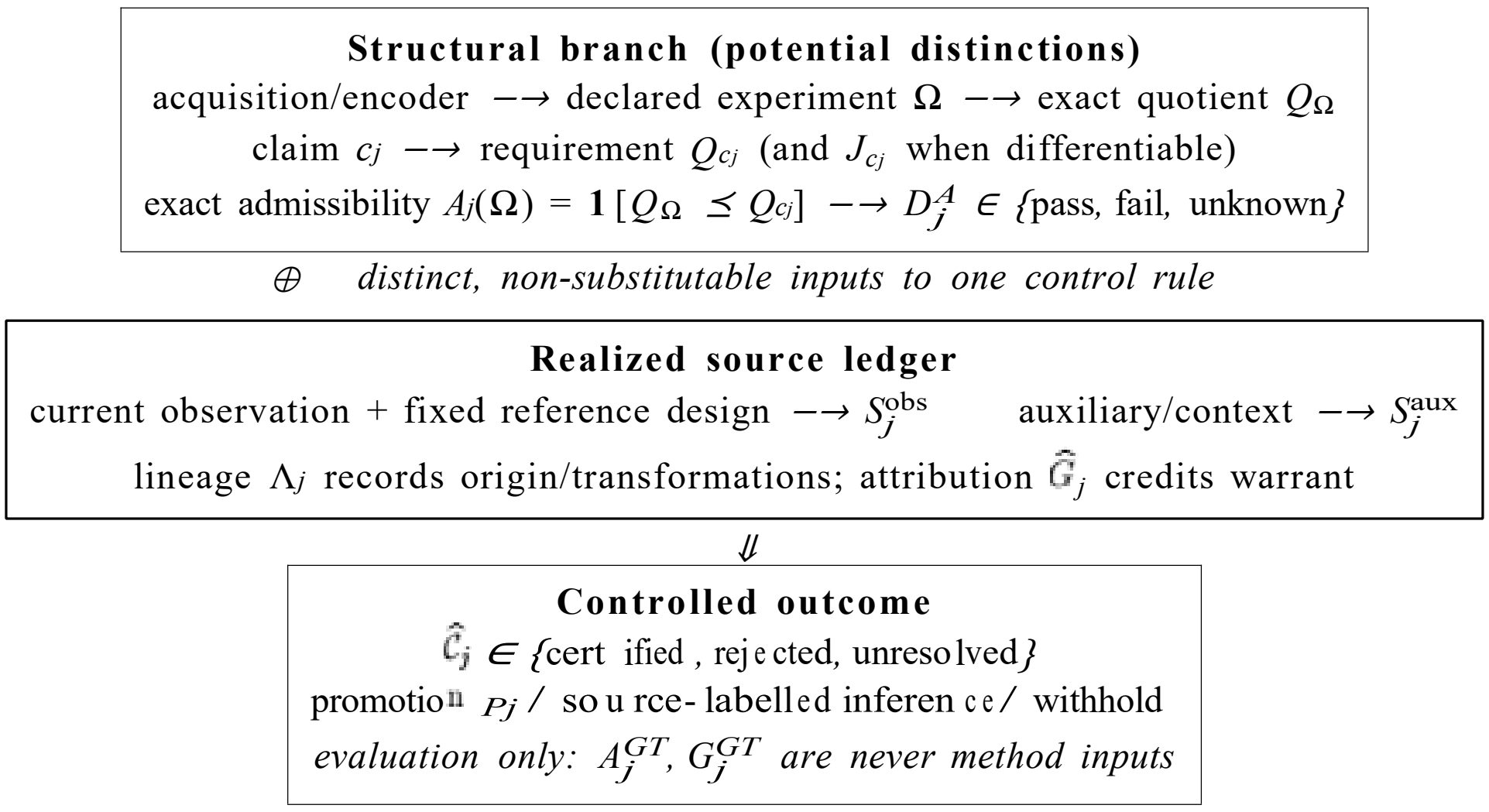

**Fig. 1 Evidence-aligned Knowledization architecture.** Acquisition and claim identity feed distinct structural and evidence/source branches. The exact quotient criterion governs structural admissibility, while realized observation and auxiliary support retain separate source semantics before certification, promotion, and routing into *K/I*. Arrows denote workflow dependency rather than logical implication; evaluator truths are used only for evaluation.

1. **Epistemic-conformance and control architecture.** We formulate a claim-level framework that keeps exact structural admissibility, observation and auxiliary support, computational lineage, epistemic source attribution, certification, and promotion formally distinct and non-substitutable. Representing these objects simultaneously enables structurally inadmissible promotion, source misattribution, unresolved-versus-rejected decisions, and source-loss/restoration to be evaluated within one interface.
2. **Exact hierarchy from classical structures.** We connect standard quotient factorization, assumption-qualified Fisher local identifiability, and classical linear estimability into an exact global/local hierarchy, while separating structural admissibility from finite-SNR operational distinguishability. The constituent mathematical principles are not claimed as new.
3. **Three-oracle evaluation.** We bind structural error to $A_j^{GT}$, realized-support error to $R_j^{GT}$, and source-relation error to $G_j^{GT}$ through IPR, RUPR, and SMR. These are operational specializations, not novel functionals.
4. **End-to-end finite execution.** We execute controlled claim proposal and grounding, learned confidence/uncertainty references, auxiliary interventions, finite *K/I* registries, four role-specific ablations, descriptive geometry diagnostics, and a typed generic-source evidence interface.

5. **Boundary studies.** High-fidelity grouping and pipeline stress tests, together with two preregistered natural-data studies, delimit the evidence: computational decomposition is not validated epistemic provenance, and physical source availability is not claim-specific realized support.

The final controlled program supports all five legacy experiments and establishes the acquisition, admissibility, semantic-separation, source-attribution, and finite-registry mechanisms at their declared controlled scope. The real-world contribution is boundary evidence rather than positive natural-data efficacy: the results identify semantics that a defensible real-world instantiation must supply.

# 2 Related Foundations

### *Structural identifiability and statistical experiments.*

Blackwell comparison, sufficiency, and statistical-experiment ordering provide classical ways to compare information through garbling [7, 15, 16]. Global and local identifiability ask which parameter distinctions are recoverable, with Fisher-rank reasoning requiring regularity and becoming delicate in singular or redundant models [8, 17, 18]. We restrict these tools to claim functions: the controller asks whether a claim is constant on observation-law fibres, rather than asserting full-state identifiability.

### *Selective prediction, abstention, and risk control.*

Reject-option classification, selective prediction, learning to defer, conformal prediction, and conformal risk control provide mature decision mechanisms for withholding or controlling error [5, 6, 13, 14, 19, 20]. Our three-way output overlaps substantially with this literature. The distinction is semantic: rejected records have a sound structural witness, unresolved records lack a justified certificate, and certified records satisfy declared structural and support conditions before promotion. Recent evidence-aware selection can also acquire additional information before answering [21]; our narrower focus is the simultaneous representation of admissibility, source-specific support, lineage, and epistemic attribution.

### *Evidence fusion, source reliability, and multimodal conflict.*

Dempster–Shafer and evidential-learning methods represent support and uncertainty, while trusted multi-view and discounted belief-fusion methods explicitly address source reliability and conflict [9, 10, 22–25]. These methods establish source awareness as a mature concern. Our narrower design requirement is non-substitution among realized claim support, exact structural admissibility, lineage, and certification.

### *Provenance and source attribution.*

Data lineage and provenance are mature database concepts, including why/where provenance and semiring accounts [11, 12]. Knowledge representation likewise tracks structured relations [26–28]. We use lineage in this mature origin-and-transformation sense, while epistemic source attribution is a separate claim-indexed output; neither object is proof of support or admissibility.

#### *Redundancy, synergy, and multimodal information decomposition.*

Partial information decomposition distinguishes unique, redundant, and synergistic information, and recent work extends these ideas to multimodal interactions [29, 30]. Our observation-only/auxiliary-only, redundant, and joint-only source classes are analogous to those categories, not a fundamentally new information taxonomy. They are instead claim-defined, intervention-controlled benchmark states, augmented with unsupported and conflicting cases, and used to evaluate promotion and source assignment rather than decompose mutual information.

#### *Claim–evidence supervision and evidence-aware selection.*

Claim-verification and rationale benchmarks such as FEVER, SciFact, and ERASER demonstrate that claim–evidence supervision is not universally absent [31–33]. Recent and concurrent work also studies evidence acquisition, claim-level specificity, certification, and fact-level multimodal attribution [21, 34–36]; the latter three are concurrent 2026 works, with the first two currently preprints. The gap exposed here is narrower: the audited multimodal physical-sensor benchmarks did not natively provide the externally grounded claim–source relation required by our source-aware evaluation.

#### *Epistemic-integrity measurement.*

A companion epistemic-integrity measurement perspective organizes Formation, Admissibility, Support, Attribution, and Lineage (FASGL) as candidate assurance roles while distinguishing the state of an individual claim from system competence, decision policy, and measurement instrument. It asks both whether a representation preserves distinctions required for downstream epistemic decisions and whether an instrument can identify its declared measurement target. This ontology and validity account is broader than the method studied here. Its additional validity conditions concern the measurement instrument rather than the controller's individual decisions. KNowLEDIzATIoN does not estimate a system-level competence profile or validate a general measurement instrument. It supplies a bounded claim-level realization: exact observation-relative admissibility, typed observation and auxiliary support, explicit source attribution and lineage, a three-state certification rule, operational promotion, and controlled evidence for their non-substitution. The relationship is therefore one of measurement perspective to criterion-referenced conformance architecture, not two names for the same contribution.

These literatures supply mature components. KNowLEDIzATIoN's contribution is to compose them into one claim-level epistemic-control interface while keeping their semantics non-substitutable, then to evaluate promotion and source attribution with independent oracles where available.

## 3 Claim-Level Epistemic-Conformance Framework

Let $X \in \mathcal{X}$ denote a latent world state. A declared acquisition configuration induces a Markov kernel

$$\Omega : \mathcal{X} \rightsquigarrow \mathcal{Z}, \qquad x \longmapsto P_\Omega(\cdot \mid x).$$

A claim is a measurable map $c_j : d' \to C_j$. The exact theory consumes a finite candidate set. Our operational pipeline instantiates that interface with a controlled five-template grammar, deterministic Top-$K$ proposal, and exact grounding; this finite experiment does not solve unrestricted claim generation or grounding.

## Finite claim formation and grounding

Let $B_C = \Phi_{\text{primitive}}(\varepsilon)$ contain declared entities, attributes, relations, and states extracted from evidence $\varepsilon$. A finite ontology and template library instantiate candidates

$$\bar{H} = \{(\bar{c}, \theta) : \bar{c} \in C_{\text{temp}},\ \ \theta \in \Theta(B_C),\ \ \text{Valid}(\bar{c}, \theta) = 1\}, \qquad H = \text{TopK}(\bar{H}; s_{\text{prop}}, K).$$

Grounding maps each proposed symbolic claim back to its declared referents and evidence records. Epistemic guarantees are conditional on the relevant claim being represented in $H$ and correctly grounded. We therefore measure candidate coverage as empirical recall over a prespecified finite target universe, separately from downstream certification conditional on inclusion. Upstream proposal or grounding errors cannot be repaired by a downstream certificate.

**Definition 3.1** (Observation equivalence). *For $x, x' \in d'$, write*

$$x \sim_\Omega x' \iff P_\Omega(\cdot \mid x) = P_\Omega(\cdot \mid x')$$

*as equality of probability measures. Let $Q_\Omega = d'/\sim_\Omega$ and $q_\Omega : d' \to Q_\Omega$ be the quotient and quotient map.*

The claim quotient $Q_{c_j}$ is induced by $x \sim_{c_j} x'$ iff $c_j(x) = c_j(x')$. We write $Q_1 \preceq Q_2$ when every class of $Q_1$ is contained in a class of $Q_2$; thus $Q_1$ is the finer partition.

**Definition 3.2** (Exact admissibility). *Claim $c_j$ is observation-admissible under $\Omega$ when*

$$x \sim_\Omega x' \implies c_j(x) = c_j(x') \qquad \text{for all } x, x'.$$

This is a structural property of the declared experiment and claim. It is distinct from the correctness of a particular predictor and from finite-sample confidence. We write

$$A_j(\Omega) = \mathbf{1}[Q_\Omega \preceq Q_{c_j}],$$

and use $A_j^{GT}$ only when a generator or proved model class supplies this exact value externally for evaluation. An internal procedure produces $D_j^A$, not $A_j^{GT}$. Equivalently, imagine collapsing every collection of states with the same observation law to one point. An admissible claim may label different collapsed points differently, but it cannot assign two values inside one collapsed point. Failure of this condition says that the declared observation process cannot establish the claim's required distinction; it does not say that the proposition is false or uninferable from every information source.

For a finite declared claim library $L$, define the exact admissible set

$$A(\Omega; L) = \{c \in L : Q_\Omega \preceq Q_c\}.$$

Let $D_j^A \in \{\text{pass, fail, unknown}\}$ denote the controller's internal structural procedure output. The structural certified-set eligibility state is

$$\overline{\text{CSET}}(\Omega; L) = \{c_j \in L : D_j^A = \text{pass}\}.$$

The two sets coincide only when the available certificate procedure is complete on the declared library. This notation is important for interpreting acquisition experiments: The acquisition intervention evaluated exact admissibility and certified-set eligibility, not every downstream condition required for operational promotion.

## 3.1 Knowledization as Claim-Level Epistemic Conformance

A broader epistemic-integrity measurement perspective distinguishes the epistemic state of an individual claim from system-level competence, the policy that maps a record to a decision, and the instrument used to assess the target. KNowLEDIzATIoN occupies the claim-level part of this perspective. Its finite candidate identity and grounding instantiate claim formation, with all downstream conclusions conditional on inclusion and correct referent binding. Admissibility then has three deliberately separate representations: $A_j(\Omega)$ is the structural property of the declared observation experiment and claim, $A_j^{GT}$ is an evaluator-only truth where an exact generator or model class supplies one, and $D_j^A$ is the controller's internal structural procedure. None may substitute for the others.

Support is likewise a typed assurance role rather than one scalar. The record retains $S_j^{\text{obs}}$ and $S_j^{\text{aux}}$ under their separate reference designs, while $R_j^{GT}$ denotes an independent evaluator relation when available. Epistemic attribution is represented by the method output $\widehat{G}_j$ and evaluated against $G_j^{GT}$; computational lineage remains the distinct transformation record $\Lambda_j$. Certification and promotion are downstream policy outputs over these declared fields, and the finite $K/I$ registries are consequences of that policy rather than additional assurance primitives.

This structured record is decision-sufficient only relative to declared decisions: it must not collapse claim states that require different certification, attribution, or promotion outcomes. That requirement explains the non-substitution contract. Equal task confidence cannot make two records epistemically equivalent when their structural, support, source, or lineage states differ. The correspondence defines a claim-level conformance realization, not a system-level latent competence measurement. In particular, the observation-model Fisher geometry used below differentiates observation laws with respect to world state; instrument-information geometry for competence measurement would concern a different likelihood and parameter space and is not evaluated here. No competence construct is inferred from this correspondence.

# 4 Theory: Exact Structure and Local Geometry

## 4.1 Global quotient characterization

Once exact admissibility is defined as constancy on observational fibres, the following factorization is the standard universal property of a quotient. Its role here is not a new quotient theorem: it is the exact global structural interface consumed by the epistemic controller.

**Theorem 4.1** (Global admissibility characterization). *For any declared experiment $\Omega$ and claim $c$,*

$$Q_\Omega \preceq Q_c \quad \Longleftrightarrow \quad c = \tilde{c} \circ q_\Omega$$

*for a unique function $\bar{c}$ on the image of $q_\Omega$.*

The elementary quotient proof is given in Appendix A.

**Proposition 4.2** (No refinement by state-independent encoding). *If an encoder is a state-independent Markov kernel $E : Z \rightsquigarrow Y$, then equality of upstream observation laws implies equality after encoding. Hence $Q_\Omega \preceq Q_{E \circ \Omega}$: encoding cannot create exact distinctions absent upstream.*

**Proposition 4.3** (Experiment refinement). *If $\Omega_1$ is a garbling of $\Omega_2$, then $Q_{\Omega_2} \preceq Q_{\Omega_1}$, and every claim admissible under $\Omega_1$ is admissible under $\Omega_2$.*

Garbling and informativeness are classical in Blackwell's comparison of experiments [7, 15, 16]. Our partition order records only equality-of-law distinctions and the induced implication for claim admissibility; it is therefore a restricted consequence of that richer decision-theoretic order, not a new information order. These statements concern exact distinctions between laws. They do not imply that every more informative experiment improves a finite-sample estimator.

## 4.2 Regularity-qualified Fisher screening

Suppose $X \subseteq$ 耽$^d$, the model is differentiable in quadratic mean, and $\Gamma_\Omega(x)$ is its Fisher information matrix. Let $F_x = [x]_\Omega$ denote the observation fibre.

The differential principle that estimable claim directions must annihilate locally unidentifiable tangent directions is classical local-identifiability reasoning [8, 17, 37, 38]. We embed that principle in the global decision hierarchy and make its regular-fibre premise explicit so that a singular diagnostic failure does not become an automatic structural rejection.

**Assumption 4.4** (Regular fibre). *Near $x$, $F_x$ is an embedded manifold and*

$$T_x F_x = \mathcal{N}(\Gamma_\Omega(x)).$$

This equality is substantive; differentiability alone does not imply it.

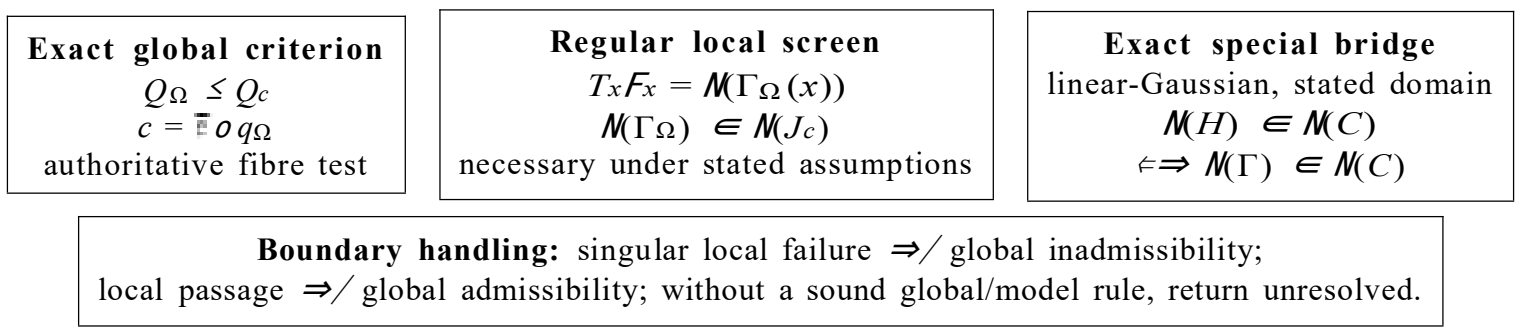

**Fig. 2 Hierarchy of global and local admissibility statements.** The quotient criterion is exact; Fisher geometry is a qualified local screen. Classical linear–Gaussian estimability supplies an exact bridge, whereas nonlinear aliasing, singular points, and finite-noise changes require explicit boundary handling. Detailed counterexamples appear in Figure 6.

**Theorem 4.5** (Local necessary condition). *Under the regular-fibre assumption, if a differentiable claim $c$ is globally admissible, then*

$$\mathcal{N}(\Gamma_\Omega(x)) \subseteq \mathcal{N}(J_c(x)), \qquad \mathrm{row}(J_c(x)) \subseteq \mathrm{range}(\Gamma_\Omega(x)).$$

*Proof* An admissible claim is constant on $F_x$, so its derivative vanishes on $T_x F_x$. Substitute the regular-fibre equality and use orthogonal-complement duality for the row-space statement. □

Outside this regular class, the null-space test is only diagnostic. For $Z \mid x \sim \mathcal{N}(x^3, 1)$, and independently for $Z \mid x \sim \mathrm{Bernoulli}(1/2 + x^3/4)$ on $(-1, 1)$, the distribution map is injective but Fisher information is zero at $x = 0$. A first-order failure therefore cannot justify structural rejection. Conversely, local passage is not sufficient globally: with $Z \mid x \sim \mathcal{N}(x^2, 1)$ on a two-sided domain, Fisher information is positive away from zero, while $x$ and $-x$ remain globally aliased. In such unverified or singular cases, the controller returns unresolved unless an exact or model-class certificate is available.

Figure 2makes the logical hierarchy explicit. The diagram does not assign certification power to a Fisher diagnostic outside its theorem domain. Figure 6consolidates the four controlled boundary cases: linear–Gaussian coincidence, cubic singularity, global aliasing, and finite-SNR strength change.

## 4.3 Exact linear-Gaussian bridge

The null-space condition below is classical linear estimability [39]. Its contribution here is connective: under the stated assumptions it makes the global quotient test, local Fisher null space, and familiar estimability criterion exactly coincide.

**Theorem 4.6** (Linear-Gaussian equivalence). *Let $\mathcal{X} = \mathbb{R}^d$ (or a nonempty open subset under the stated linear-domain conditions),*

$$Z = HX + \varepsilon, \qquad \varepsilon \sim \mathcal{N}(0, \Sigma), \qquad \Sigma \succ 0,$$

*and $c(x) = Cx$. Then*

$$Q_\Omega \preceq Q_c \quad \Longleftrightarrow \quad \mathcal{N}(H) \subseteq \mathcal{N}(C) \quad \Longleftrightarrow \quad \mathcal{N}(\Gamma) \subseteq \mathcal{N}(C),$$

*where* $\Gamma = H^T\Sigma^{-1}H$.

The short estimability proof is given in Appendix A.

For arbitrary restricted domains, the pairwise fibre rule remains authoritative; the full null-space necessity need not follow when some null directions are not realized as domain differences.

### 4.4 Structural versus operational distinguishability

With fixed $H$ and every finite positive noise scale $\sigma$, the linear-Gaussian quotient is determined by $Hx$, not by the magnitude of $\sigma$. Yet

$$\Gamma_\sigma = \sigma^{-2}H^TH$$

and finite-sample total-variation separation between distinct laws decreases with noise. Exact admissibility and operational distinguishability therefore answer different questions. We use Fisher spectra and total variation descriptively; coordinate-dependent geometry scores are heuristics, not certificates.

### 4.5 Descriptive spectral and directional geometry

For a positive-semidefinite Fisher matrix $\Gamma = V\,\mathrm{diag}(\lambda_k)V^T$, the completed diagnostic program reports algebraic rank, effective rank

$$r_{\mathrm{ef}} = \#\{k : \lambda_k/\lambda_{\max} > 10^{-3}\},$$

with $r_{\mathrm{ef}} = 0$ when $\lambda_{\max} = 0$, stabilized $\log\det(\Gamma + 10^{-6}I)$, and directional information $d^T\Gamma d$. The claim-geometry mismatch diagnostic uses

$$P_{\mathrm{weak}} = V\,\mathrm{diag}\left(\frac{0.05}{\lambda_k + 0.05}\right)V^\top, \qquad \mathrm{CGM}\,(c,\Gamma) = \frac{\|J_cP_{\mathrm{weak}}\|_F}{\|J_c\|_F + 10^{-12}}.$$

These quantities depend on the declared coordinate/scale conventions. None supplies support, attribution, rejection, certification, or promotion authority.

## 5 Source-Aware Epistemic Control

### 5.1 A definition-complete interface

Appendix Table 8 fixes the mathematical roles used by the controller and evaluation. These are semantic interfaces, not programming-language types. In particular, evaluator-only oracles are joined only for evaluation and are never method inputs.

Admissibility asks what the experiment could establish. Realized support asks what the current record supplies for claim $j$ under a declared source/reference design. For a binary claim hypothesis $h \in \{+, -\}$, one observation design supplies measures $\mu_h$ and

$$p_{\mathrm{obs}}(z \mid h) = \int p_\Omega(z \mid x)\, d\mu_h(x), \qquad S_j^{\mathrm{obs}}(z) = \log \frac{p_{\mathrm{obs}}(z \mid +)}{p_{\mathrm{obs}}(z \mid -)}.$$

Thus $S_j^{\mathrm{obs}}$ is record-, claim-, channel-, and reference-design-relative; it is not prior-free and is not defined by confidence, channel presence, or a neural branch.

For an auxiliary record $a \in A_j$ and declared auxiliary design $D_j^{\mathrm{aux}}$, define separately

$$S_j^{\mathrm{aux}} = s_j^{\mathrm{aux}}(a; D_j^{\mathrm{aux}}) \in V_j^{\mathrm{aux}}.$$

The codomain may be graded or discrete. In the synthetic studies it is a prespecified claim-relative log-support component; in the natural-data studies a comparable claim-specific auxiliary value was unavailable unless explicitly operationalized. $S_j^{\mathrm{aux}}$ neither inherits observation semantics nor alters $A_j(\Omega)$. More generally, $S_j^{\mathrm{obs}}$ and $S_j^{\mathrm{aux}}$ need not be numerically comparable. They are additive only under an explicitly declared joint model; conditional independence before latent mixing or architectural separation is insufficient [9, 10, 25].

Lineage $\Lambda_j$ answers where data or computation originated. Epistemic attribution $\hat{G}_j$ answers which source relation the controller credits with warrant. Mature database provenance already formalizes lineage [11, 12]; our claim-indexed ledger retains it but does not promote it into support. Computational lineage and epistemic attribution remain separate even when an implementation serializes them in one record. Hence

$$\boxed{\text{lineage} \not\Rightarrow \text{epistemic support}}, \qquad \boxed{\text{computational contribution} \not\Rightarrow \text{epistemic attribution}}.$$

For controlled source evaluation,

$$G = \{\text{obs, aux, redundant, joint, unsupported, conflicting}\}.$$

$G_j^{GT}$ is computed before method output by prespecified observation and auxiliary neutralizations: individual support above the declared threshold gives observation-only or auxiliary-only; both give redundant; neither individually but the combination gives joint-only; neither gives unsupported; and opposed sufficiently strong signs give conflicting with precedence. This source-class oracle can be multi-source and is external to the controller. A method prediction is correct only when $\hat{G}_j = G_j^{GT}$; unknown counts as error. Natural data do not acquire $G_j^{GT}$ from branch identity or source presence.

The three external truths answer different questions:

$$A_j^{GT} = \text{structural admissibility}, \qquad R_j^{GT} = \text{realized support}, \qquad G_j^{GT} = \text{source relation}.$$

When the controlled generator exposes a continuous observation likelihood ratio $L_{O,j}^{GT}$, it remains distinct from the binary policy label $R_j^{GT}$ and from the method estimate $\hat{L}_{0,j}$. Oracle quantities may supervise declared training records, but none is available to the inference-time controller.

## 5.2 Typed evidence access and finite registries

The executable generic-source GET interface represents evidence, source identity, acquisition, grounding/referent links, coordinate systems, Fisher geometry, support, uncertainty, lineage, transformations, attribution, grouping, dependence/fusion state, and a derived evidence ledger. “Unified” means common typed access, not numerical interchangeability of modalities. The controller adapter preserves these types and does not itself compute admissibility, certification, or promotion. A positive ledger entry is therefore not a certificate.

For a finite evaluated claim universe at stage $t$, define the operational promoted registry and auxiliary-supported inference registry

$$K_t = \{j : P_{j,t} = 1\}, \qquad I_t = \{j : \text{the declared auxiliary-routing contract passes and } j \notin K_t\}.$$

These are protocol-relative finite registries, not universal stores of knowledge. Their state is distinct from the three-state certification object.

## 5.3 Support gate, certification, and promotion

Let declared source gates be $g_{s,j} : V_j^s \to \{\text{pass, fail, unknown}\}$ for $s \in \{\text{obs, aux}\}$. A gate is part of the declared claim contract; it is not learned from outcome performance. In the controlled B5 studies, $g_{\text{obs},j}$ = pass iff $S_j^{\text{obs}} > 1$ and is fail otherwise. The observation-certified policy does not let auxiliary support substitute for this gate.

Let $V_j \in \{\text{pass, fail, unknown}\}$ summarize the required, declared source-relation checks. The final certification mapping is

$$\widehat{C}_j = \kappa(D_j^A, g_{\text{obs},j}, g_{\text{aux},j}, \Lambda_j, \widehat{G}_j) = \begin{cases} \text{rejected}, & D_j^A = \text{fail}, \\ \text{certified}, & D_j^A = \text{pass},\ g_{\text{obs},j} = \text{pass},\ V_j = \text{pass}, \\ \text{unresolved}, & \text{otherwise.} \end{cases}$$

This maps the two-stage implementation—structural certificate followed by the support eligibility gate—into one explicit interface. A rejection is a sound structural witness, not a false task label. Unknown structure, unknown support, or an unverified required source relation is unresolved rather than rejected. The mapping is an architectural contract; restricted real-world studies instantiate only their declared components. In the controlled B5 experiments, $V_j$ = pass by the prespecified contract: source attribution was recorded and evaluated but was not an additional promotion gate.

For the controlled B5 promotion policy,

$$P_j = \mathbf{1}[\widehat{C}_j = \text{certified}].$$

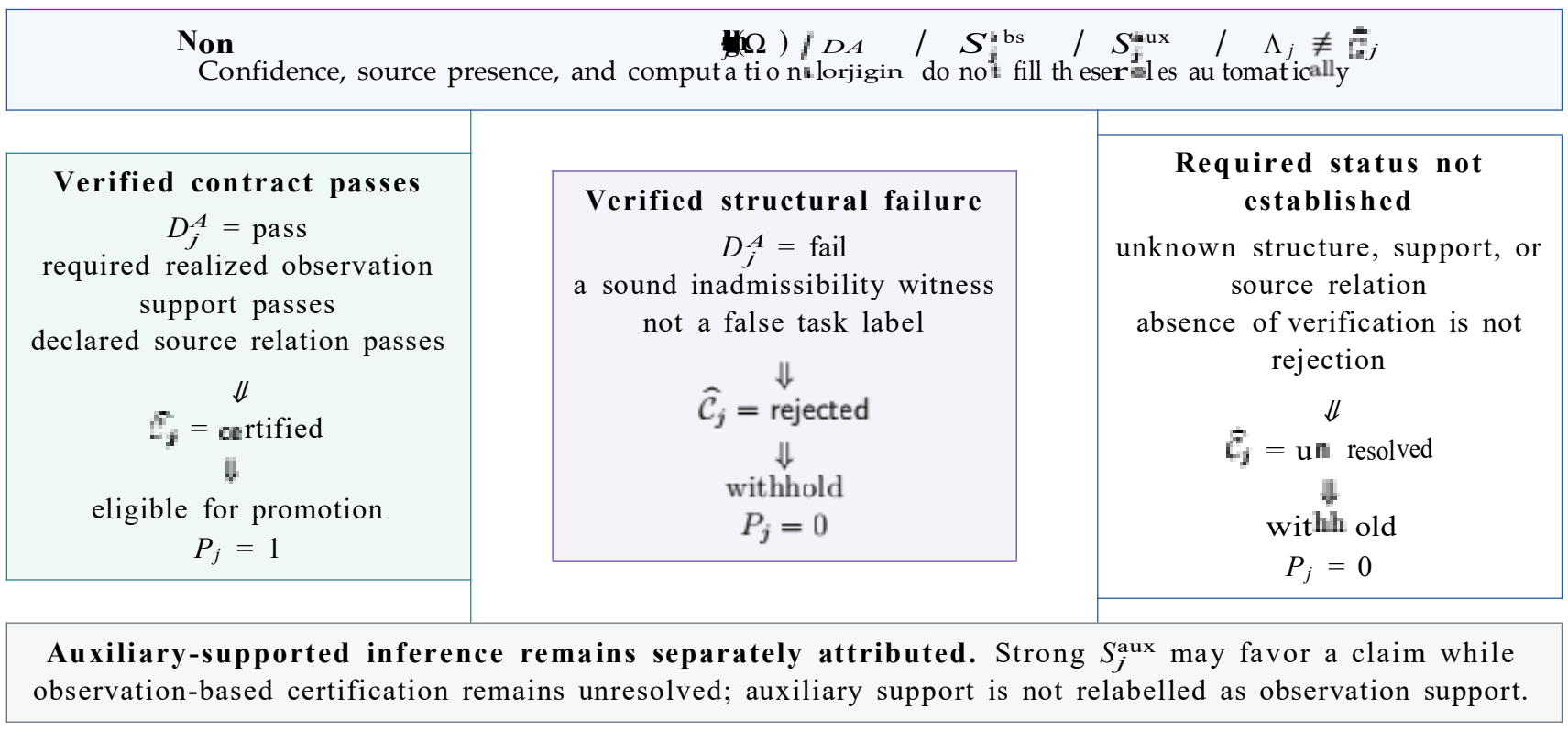

**Fig. 3 Detailed semantic controller states.** The declared ledger keeps support, attribution, structural decision, certification, and promotion distinct. Verified structural failure produces rejection, an unverified required condition produces an unresolved state, and only the complete passing contract permits promotion. Auxiliary support may sustain separately attributed inference but cannot become observation support.

Thus rejected and unresolved claims are not promoted, while a structurally passing claim may still be withheld by the support/source gate. Promotion is an operational release decision, not philosophical knowledge.

### *Semantic non-substitution contract.*

Absent an additional declared assumption, the architecture forbids

$$\text{confidence} \Rightarrow \text{support}, \qquad \text{lineage} \Rightarrow \text{epistemic attribution},$$
$$\text{source availability} \Rightarrow \text{realized support}, \qquad \text{support} \Rightarrow \text{structural admissibility},$$
$$\text{local Fisher passage} \Rightarrow \text{global admissibility}, \qquad \text{geometry fusion} \Rightarrow \text{support fusion},$$
$$\text{ledger positive} \Rightarrow \text{certification}, \qquad \text{support alone} \Rightarrow \text{promotion},$$
$$\text{task correctness} \Rightarrow \text{epistemic correctness}.$$

This is a design contract, not a universal theorem. Representing the objects separately makes four consequences evaluable: structural promotion error, source-attribution error, rejected-versus-unresolved behavior, and admissibility/CSET changes under acquisition loss without equating loss with claim falsity.

Figure 3 makes the three-state control semantics explicit. It shows why verified structural failure, inability to establish a required condition, and auxiliary-only support produce different records even though none is promoted under the controlled policy.

External evaluation uses independent oracles where available. The structural metric is

$$\mathrm{IPR} = \frac{\sum_j \mathbf{1}[P_j = 1, A_j^{GT} = 0]}{\sum_j \mathbf{1}[P_j = 1]},$$

the *Inadmissible Promotion Rate*. It is a specialization of promoted-set selective error using a structural-admissibility oracle; no realized-support oracle enters it. Realized

Unsupported Promotion Rate is

$$\mathrm{RUPR} = \frac{\sum_j \mathbf{1}[P_j = 1, R_j^{GT} = 0]}{\sum_j \mathbf{1}[P_j = 1]},$$

which uses only the realized-support oracle. Strict source-misattribution rate is

$$\mathrm{SMR} = \frac{\sum_j \mathbf{1}[P_j = 1, \widehat{G}_j \neq G_j^{GT}]}{\sum_j \mathbf{1}[P_j = 1]}.$$

It is a promoted-set attribution error, not a novel statistical functional. Neither metric is meaningful without its corresponding external oracle. Legacy Experiment 3 additionally reports $\mathrm{UPR}_{\mathrm{old}} = \#\{j : P_j = 1, C_j \neq \text{certified}\}/\#\{j : P_j = 1\}$. This internal-certification quantity is conceptually related to, but not mathematically identical with, external-oracle IPR; neither is RUPR.

# 6 Experimental Strategy

## 6.1 Scientific sequence and experimental safeguards

The experiments form one causal sequence rather than reducing epistemic control to a single end-to-end accuracy number. We first perturb observation structure, then ask which claim distinctions remain admissible, whether confidence and realized support substitute for structural status, how auxiliary evidence changes belief and attribution, and finally how these effects propagate into finite registries. The design distinguishes three paths:

$$\text{structural change} \longrightarrow Q_\Omega \longrightarrow \text{admissibility} \longrightarrow \text{certification eligibility},$$
$$\text{finite-noise/SNR change} \longrightarrow \text{Fisher magnitude/operational separation},$$
$$\text{source-evidence relation} \longrightarrow (S_{\mathrm{obs}}, S_{\mathrm{aux}}, \widehat{G}) \longrightarrow \text{certification/promotion}.$$

Viewed through the claim-level conformance perspective, these interventions isolate distinct assurance roles without changing their original hypotheses or endpoints. The second path does not automatically alter exact admissibility, and the third does not permit favorable evidence or confidence to substitute for structural status.

The controlled program used prespecified finite claim libraries and evaluator-side structural and source truths that were unavailable to the methods. Predictions were fixed before these truths were joined, and primary summaries were independently recomputed. This separation is necessary because ordinary task labels do not specify whether a claim is constant on observation fibres or which source relation warrants it.

Claim-formation, GET, and natural-data studies extend this controlled sequence without redefining its scientific criteria.

## 6.2 Claim proposal and final predictive interface

The claim-formation study uses six true referents per scene, a five-template controlled grammar, exact candidate enumeration, a prespecified proposal score, deterministic Top-$K$, and an evaluation-only grounding oracle. Recall of hypotheses at $K$ (RH@$K$) measures whether a target survives proposal and selection; Grounding Accuracy measures exact referent recovery among proposed claims. Both are finite empirical endpoints.

The final method attaches task confidence to each canonical record and estimates claim-conditioned auxiliary support and source attribution without changing the task predictor. Raw confidence, temperature scaling, MC dropout, a five-member ensemble, and an exact posterior serve as predictive comparators. Training supervision is confined to the declared development populations, and held-out predictions are evaluated independently of evaluator truth.

## 6.3 Structural admissibility and local/global boundaries

The first question is which acquisition interventions change exact distinctions and which change only operational evidence strength. Structural rank/channel changes modify the linear observation map $H$ and can coarsen $Q_\Omega$; a strength-only study holds $H$ fixed while varying finite Gaussian noise. The second question is when local geometry is exact, when it is only a qualified screen, and when it is insufficient. We therefore combine (i) the prespecified linear–Gaussian special case, for which quotient, null-space, and row-space criteria coincide; (ii) cubic-normal and cubic-Bernoulli singular cases, where an unverified local path must return unresolved rather than reject; and (iii) a two-sided $x^2$ observation map, where local passage can coexist with global aliasing. Figure 2 and Appendix Fig. 6 summarize these cases. The external $A_j^{GT}$ table is used only to score structural decisions; methods cannot access it.

## 6.4 Source-aware support and certification

The source experiments retain two complementary epistemic failures. In the first, a claim can have strong auxiliary or predictive evidence while the declared observation channel does not establish the required observation-grounded status. In the second, realized evidence can be favorable while structural admissibility or a sound certificate is absent. These cases are instantiated through the separate objects $(A_j, S_{\mathrm{obs},j}, S_{\mathrm{aux},j}, \hat{C}_j, P_j)$ rather than a combined confidence score. Hidden-verification controls test that lack of a sound structural procedure yields *unresolved*, not forced rejection; inadmissible high-support cells test that support cannot manufacture admissibility. Figure 3 shows the decision paths.

The promotion-safety endpoint is IPR, evaluated with structural-admissibility truth $A_j^{GT}$, not a realized-support oracle. B5’s structural eligibility makes zero IPR an expected safety invariant; the comparison asks whether that invariant survives finite estimation, ranking, deterministic tie handling, and matched coverage.

## 6.5 Source attribution under controlled interventions

The source-attribution question fixes the declared observation design and asks whether auxiliary intervention can change auxiliary evidence without being relabelled as observation support. The generator defines six counterfactual source states—observation-only, auxiliary-only, redundant, joint-only, unsupported, and conflicting—in an external $G_j^{GT}$ table. Neither the state label nor its counterfactual bits are method inputs. Under the prespecified construction, auxiliary neutralization must leave $S_{\text{obs},j}$ unchanged, and observation neutralization must leave $S_{\text{aux},j}$ unchanged. A symmetric squared-observation/sign-claim case supplies a negative control for false observation support.

The factorial construction identifies these mechanisms but makes source classes directly recoverable from factual patterns. The comparative centerpiece therefore uses overlapping, correlated evidence that entangles those patterns. It compares B5 with the information-matched strongest-source rule B-SRC using strict promoted-set SMR against the external six-state oracle. The final source-evidence readout predicts the same ontology. For Experiment 3, the common $q$ = .20 confidence universe is formed first, promotion then uses observation support plus structural certification only, and attribution is evaluated on the fixed promoted set.

## 6.6 Acquisition contraction and recovery

The longitudinal intervention is $H_0 \rightarrow H_1 \rightarrow H_2 \rightarrow H_0^{\text{restore}}$. The first two degradations remove observable directions while holding the claim family fixed; the final condition restores the original observation operator. This contrast tests whether lost distinctions withdraw only the claims that require them and whether restoring acquisition recovers the same admissibility and certification states. The finite-registry study then carries those states through the unchanged support and source-routing rules to measure the finite registries $K_t$ and $I_t$ at the same four acquisition conditions.

## 6.7 Comparison and boundary logic

Promotion-safety and source-attribution comparisons use matched coverage so that a method cannot appear safer merely by withholding more claims. Comparator information sets, deterministic selection, exact dataset construction, and inferential procedures are given in Section 9. The real-data studies extend the controlled program without repeating it: TbV asks whether exact computational branch contributions carry the required epistemic source meaning, whereas nuScenes asks whether physical direct-range source availability supplies claim-specific support. These are prospective proxy-validity tests, not positive deployment evaluations.

# 7 Results

The experiments test a causal account of claim-level epistemic qualification. Claims must first be represented and grounded. Acquisition then determines which distinctions remain observable; those distinctions constrain claim admissibility; and realized

**Table 1** Controlled experimental progression from acquisition to finite epistemic state. The rows form one mechanistic chain rather than independent status checks.

| Question / mechanism | Intervention / contrast | Primary observation | Scientific interpretation | Scope |
|---|---|---|---|---|
| Acquisition → distinctions (Exp. 1) | Remove/restore directions; vary noise | Rank 3 → 2 → 1 → 3; selective information loss | Acquisition changes observable distinctions | Controlled |
| Claim admissibility (Exp. 2) | Compare retained vs lost directions | Claims 12 → 6 → 3 → 12 | Admissibility is claim-dependent | Controlled |
| Non-substitution (Exp. 3) | Common confidence set; Cases A/B | UPR 0; SMR 0.174 → 0.015; Case B 1,331/7,500 | Confidence, support, structure, and source difer | Controlled; Case B enriched |
| Source separation (Exp. 4) | Vary auxiliary evidence only | Status invariant; belief +0.536; source error falls | Auxiliary evidence updates belief and provenance | Controlled |
| Registry response (Exp. 5) | Degrade and restore acquisition | $K$: 1381, 613, 271, 1381; $I$: 1740, 927, 554, 1740 | Both registries contract and recover | Controlled |

*Scope note.* Downstream epistemic conclusions are conditional on the relevant claim being represented in the candidate family and correctly grounded.

support and source semantics determine how qualified claims enter finite epistemic state. Table 1 summarizes this progression and the scope of each contrast.

## 7.1 Candidate formation bounds downstream epistemic conclusions

Before asking whether a claim is epistemically qualified, the relevant claim must be present and correctly grounded. The five-template grammar expressed all 177,760 controlled targets, but exact candidates were generated for 151,865, Top-32 retained 108,700, and 30,749 were exactly grounded (Figure 4). Scene-macro RH@32 was 0.664 (95% CI [0.660, 0.668]), whereas Grounding Accuracy conditional on a proposal was 0.173 (95% CI [0.172, 0.175]). The ontology could therefore express the targets, but proposal and retrieval did not always place them in the evaluated candidate set, and exact grounding was the dominant bottleneck. Downstream certification is conditional on candidate inclusion and correct grounding; it cannot repair an incorrectly grounded claim.

The complete candidate-budget curve was scene-macro RH@8 0.201071, RH@16 0.388022, RH@32 0.664067 (95% CI [0.659871, 0.668291]), and RH@64 0.842854; the corresponding pooled values were 0.176975, 0.344464, 0.611499, and 0.835548. Grounding Accuracy conditional on proposal was 0.173274 (95% CI [0.171865, 0.174725]). These secondary budgets show a progressive retrieval gain but do not remove the distinct grounding bottleneck. Equivalently, 25,895 expressible targets lacked an exact generated candidate, and the Top-32 restriction removed a further 43,165 otherwise

**Table 2** Held-out claim-generation endpoints. Only RH@32 was primary.

| Endpoint | Scene macro | Pooled micro | 95% CI |
|---|---|---|---|
| RH@8 | 0.201071 | 0.176975 | secondary |
| RH@16 | 0.388022 | 0.344464 | secondary |
| RH@32 | 0.664067 | 0.611499 | [0.659871,0.668291] |
| RH@64 | 0.842854 | 0.835548 | secondary |
| Grounding Accuracy \| proposed | 0.173274 | — | [0.171865,0.174725] |

recoverable targets. These losses and exact grounding are distinct upstream failure modes.

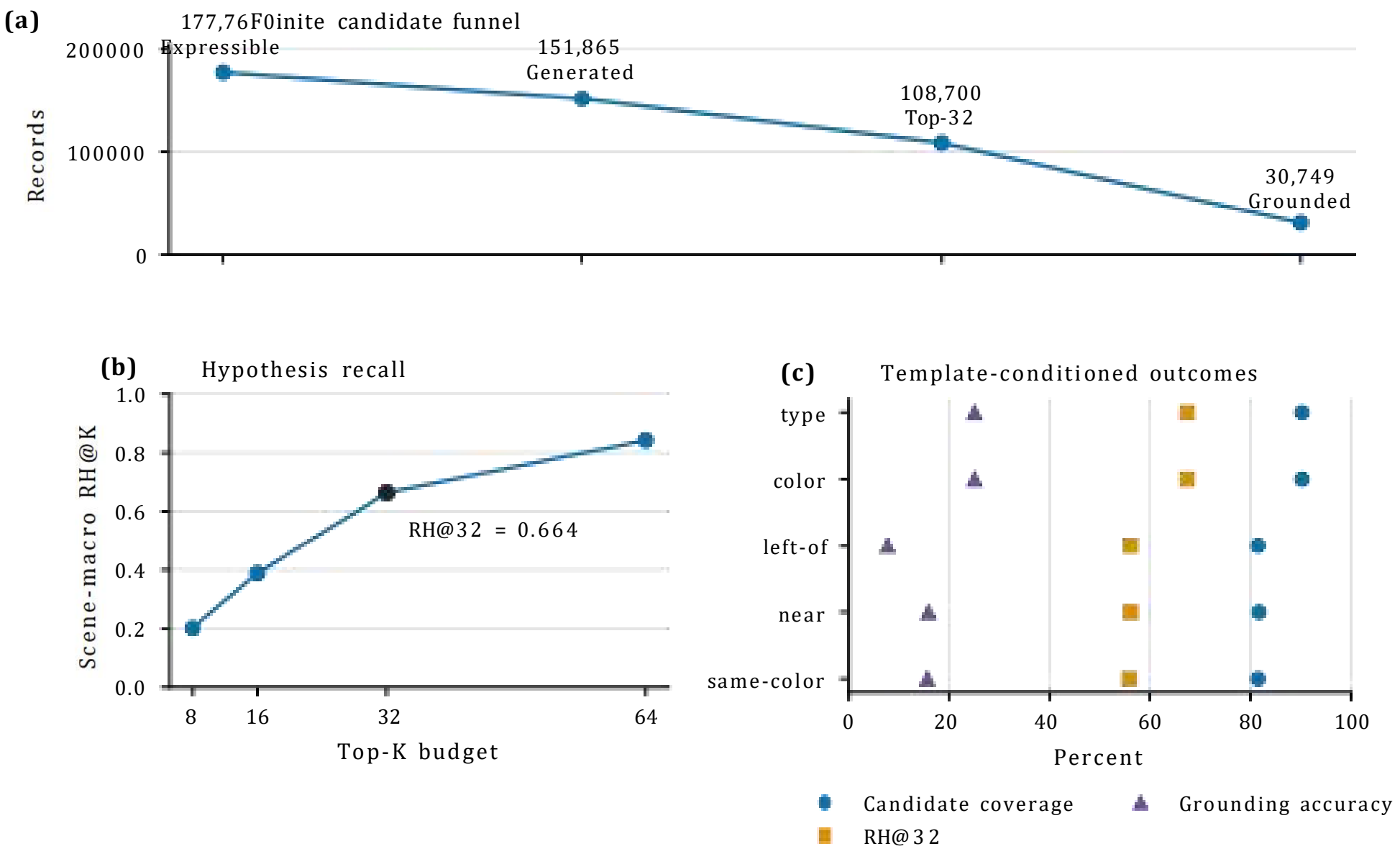

**Fig. 4** **Finite claim formation exposes a grounding bottleneck.** A fixed five-template proposal interface progressively restricts the expressible target universe (a); RH@$K$ quantifies retrieval under the declared budget (b), and template-conditioned measurements separate candidate coverage from exact grounding (c). Complete expressibility therefore does not imply proposal inclusion or correct grounding.

## 7.2 Acquisition defines the boundary of observable claims

### 7.2.1 Acquisition reshapes observable distinctions

The first mechanistic question is whether acquisition changes the distinctions available to an observer (Experiment 1). Across $H_0 \rightarrow H_1 \rightarrow H_2 \rightarrow H_0^{\text{restore}}$, the intervention removed and then restored coordinate directions while leaving the latent claim family unchanged. Effective rank contracted as $3 \to 2 \to 1$ and returned to 3; stabilized logdet and directional information changed concordantly (Figure 5). Information vanished only along removed directions and returned when acquisition was restored. By contrast, holding $H$ fixed while varying $\sigma \in \{.25, .5, 1, 2, 4\}$ changed Fisher strength through $\Gamma(\sigma) = \sigma^{-2}\Gamma(1)$ without changing the exact admissible set. These contrasts identify acquisition as a determinant of observable distinction structure, not merely a uniform perturbation of predictive confidence.

### 7.2.2 Observable distinctions constrain claims selectively

The corresponding claim-level question is whether every claim responds uniformly or only claims that require the lost distinctions do so (Experiment 2). Admissible templates contracted as $12 \to 6 \to 3$ and returned to 12: claims requiring a removed direction became inadmissible, whereas claims constant along that direction remained admissible. Thus the controlled chain was $\Delta a \to \Delta\Gamma_\Omega \to \Delta A(\Omega)$, with restoration recovering the original states. The controls in Figure 6 explain why this conclusion requires global as well as local analysis. For $Z \mid x \sim N(x^3, 1)$ at $x = 0$, local Fisher vanishes although the map is globally injective. Conversely, for $Z \mid x \sim N(x^2, 1)$ at the prespecified regular point, local Fisher is nondegenerate although $x$ and $-x$ are globally aliased. Removing global analysis falsely certified both aliased claims, whereas exact global analysis certified neither. Local Fisher geometry is therefore a qualified diagnostic of operational distinction strength: neither local weakness nor local passage establishes global inadmissibility or admissibility. These controls demonstrate non-equivalence, not the prevalence of either pathology in applications. Exact quotient admissibility therefore defines the structural boundary, while local Fisher geometry remains a qualified diagnostic outside the model classes in which the two coincide.

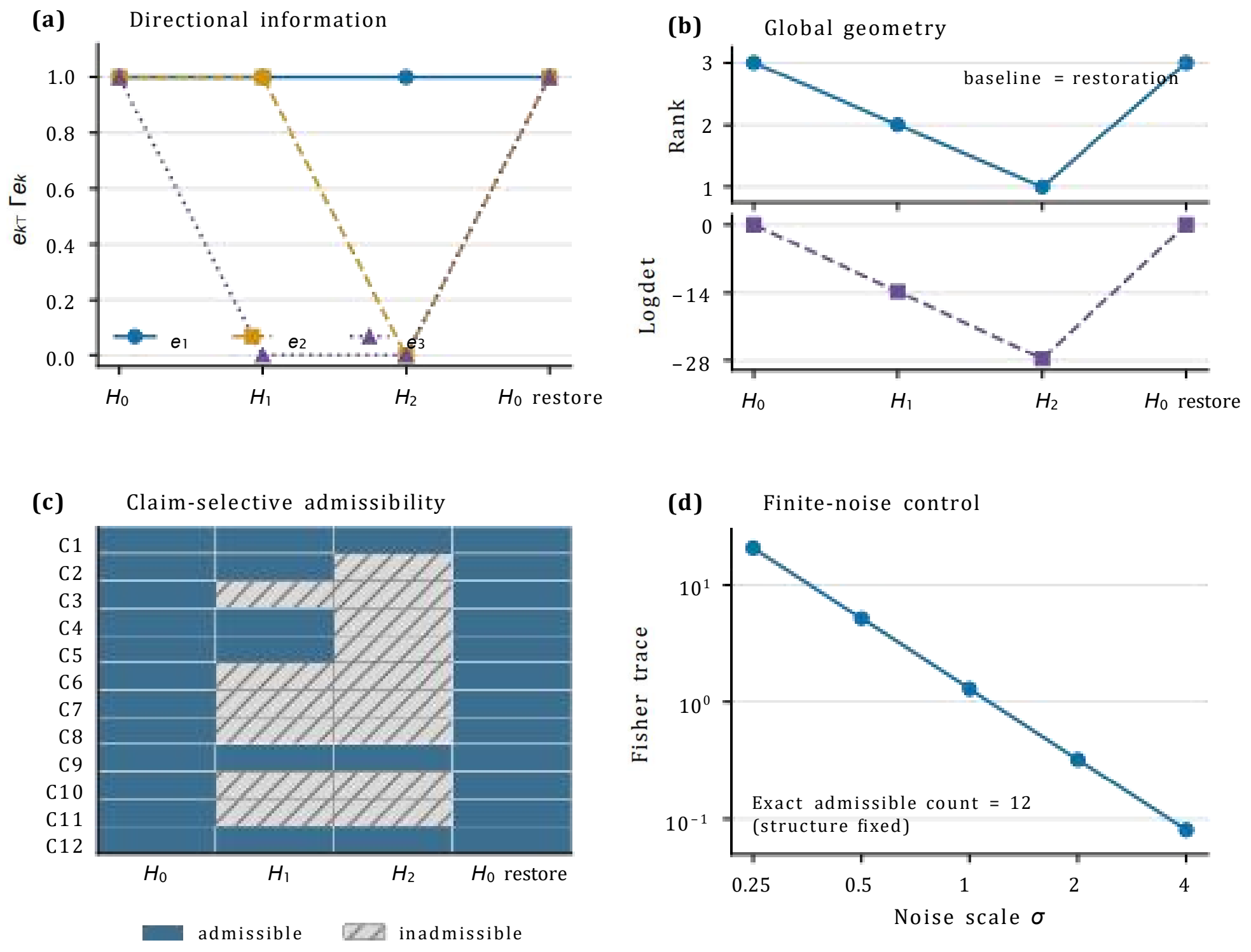

**Fig. 5 Acquisition changes observable distinctions and, consequently, claim-selective admissibility.** Removing observation directions causes the corresponding directional information, efective rank, and stabilized logdet to contract (a,b). Claims requiring the lost directions become inadmissible while others remain available (c), and restoration reproduces the baseline structure. Varying finite noise at fixed structure changes Fisher strength but leaves the exact admissible count unchanged (d).

#### 7.2.3 Global/local and finite-sample diagnostic boundaries

The analytic boundary cases make the logical status of the geometric diagnostics explicit (Figure 6). The null-space rule is exact in the stated linear–Gaussian class. By contrast, the cubic-normal and cubic-Bernoulli maps are globally injective even though their first-order Fisher information vanishes at the prespecified singular point, while the two-sided $x^2$ map passes a local screen away from zero despite global aliasing. Varying finite noise at fixed structural map changes Fisher magnitude and finite-sample separation without changing exact observation fibres.

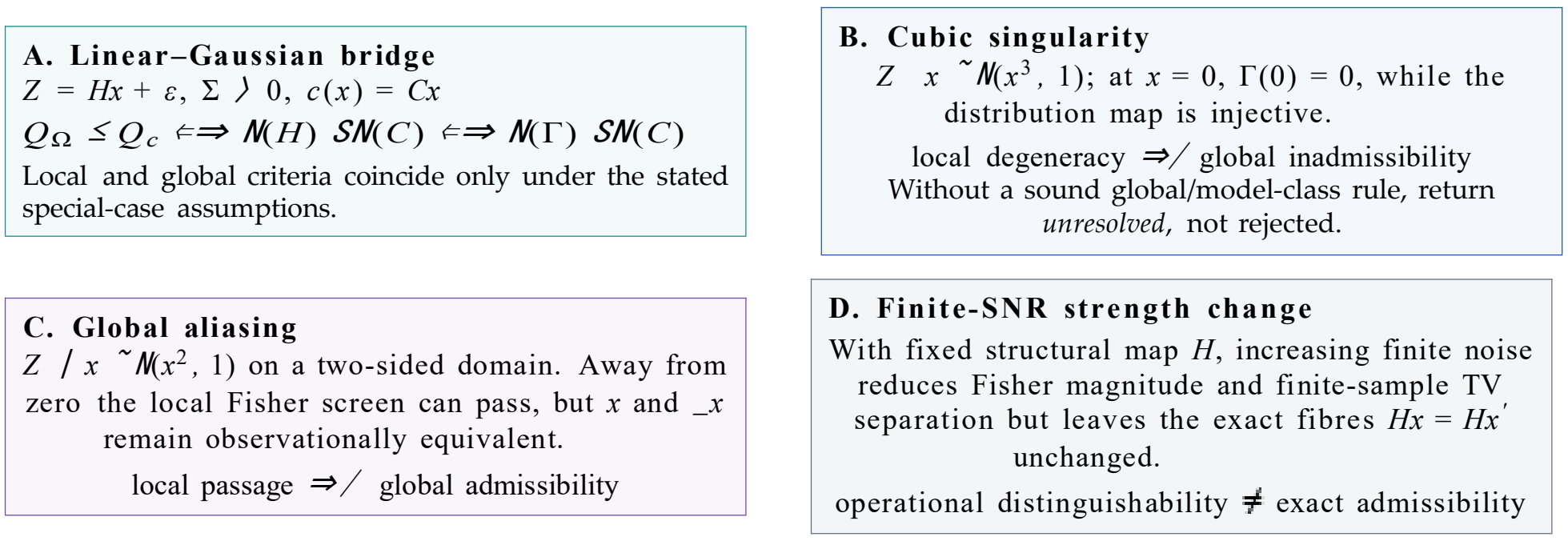

**Fig. 6 Where global and local structural tools agree or fail. A,** the null-space criterion is an exact classical bridge in the stated linear–Gaussian domain. **B,** a singular first-order diagnostic can be uninformative even when the distribution map is globally injective. **C,** a local screen can pass while a two-sided observation map remains globally aliased. **D,** finite noise changes Fisher magnitude and operational separation without changing the exact observation fibres when the structural map is fixed. Panels B and C therefore return unresolved unless a sound global or model-class certificate is available; they do not license an unconditional Fisher rejection rule.

The frozen degeneracy family $\Gamma(\lambda) = \mathrm{diag}(1, .3, \lambda)$ separately tests the behavior of descriptive geometry as one direction weakens (Figure 7). Algebraic rank remained three for every positive $\lambda$ and became two at zero; under the strict relative threshold, effective rank remained three through $\lambda = .003$, became two at $\lambda = .001$, and remained two at the separate zero boundary. Stabilized logdet decreased from $-1.203967$ at $\lambda = 1$ to $-8.110724$ at $\lambda = .001$ and $-15.019479$ at zero. Directional information along $e_3$ tracked $\lambda$, while its CGM increased from 0.047619 at $\lambda = 1$ to 0.980392 at $\lambda = .001$ and 1.000000 at zero. The weakest population space was $\mathrm{span}(e_2)$ for $\lambda > .3$, $\mathrm{span}(e_2, e_3)$ at the exact crossing $\lambda = .3$, and $\mathrm{span}(e_3)$ below it.

Held-out mean weak-subspace error decreased from 0.115761 at $N = 128$ to 0.100522 at $N = 512$ and 0.090366 at $N = 2048$, but the exact eigenspace crossing at $\lambda = .3$ retained elevated error. The effective-rank/weak-error aggregate was defined at $N = 128$ but was N/A at $N = 512$ and $N = 2048$, whose defined-seed fractions 0.25 and 0.35 were below the declared 0.80 requirement. All unresolved-rate rank correlations were N/A because every estimator status was valid. These null and non-evaluable outcomes are part of the result: spectral, directional, logdet, and CGM measurements diagnose finite-sample geometry but do not certify global admissibility.

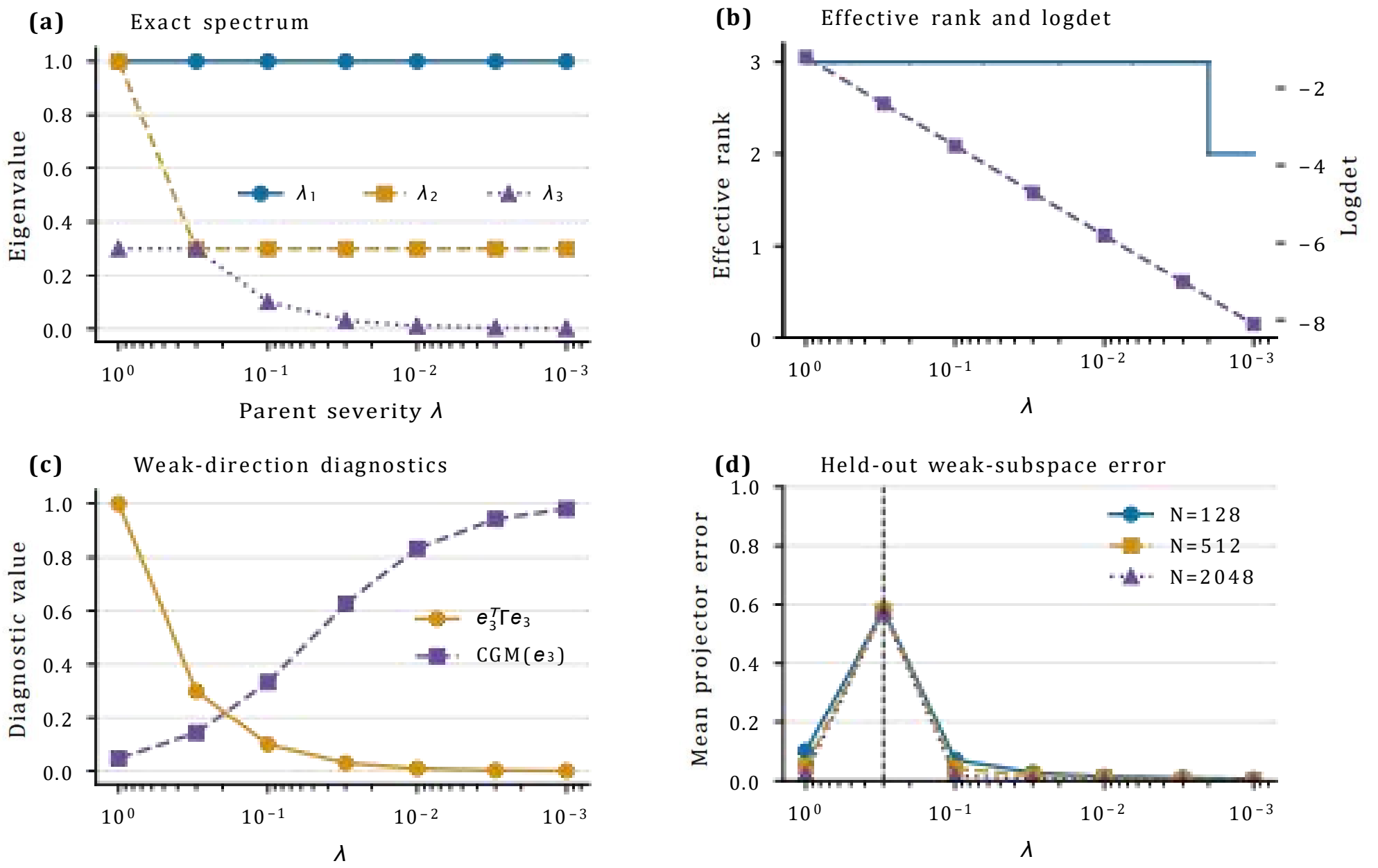

**Fig. 7 Controlled degeneracy diagnostics (diagnostic only).** The parent family $\Gamma(\lambda) = \mathrm{diag}(1, .3, \lambda)$ weakens a declared direction over a prespecified grid (a–c). Finite-sample held-out projector error is shown for the three sample sizes (d); the peak at $\lambda = .3$ is the exact population eigenspace crossing, not an implementation defect. Effective rank, logdet, directional information, and CGM are descriptive and do not certify global admissibility.

## 7.3 Confidence does not determine epistemic qualification

Predictive confidence is often treated as a proxy for reliability. Experiment 3 asks whether a scalar confidence score preserves the distinctions required for epistemic qualification. Case A tests whether high confidence can coexist with insufficient observation-relative support; Case B tests whether strong instance-level support can override structural non-identifiability. In the general held-out population, 6,000 H0-neutral records produced a deterministic $q = .20$ common confidence universe of 1,200. Observation support plus structural certification promoted 477 claims. Historical UPR—the promoted fraction whose internal certification was not certified—was 0, while Case A occurred 1,854 times in the full H0-neutral population. Confidence therefore supplied neither observation support nor structural certification.

Source identity provided a further distinction that confidence could not recover. On the same promoted records, the source-aware readout reduced promoted-set SMR from 0.174 to 0.015 (Δ = __0.159, paired seed-cluster 95% CI [__0.205, __0.135]). Attribution accuracy, balanced accuracy, and macro-F1 are reported in Table 3. Source attribution is therefore a separate epistemic role, not a relabeling of predictive confidence.

The general random population contained no Case-B record. To test that mechanism directly, Case B was examined in a separate prospectively enriched controlled population designed to instantiate the prespecified $H_1/H_2$ structural contrast. Among 7,500 design-side opportunities, all 7,500 remained structurally rejected while 1,331 had $\hat{L}_O > 1$, yielding 1,331 joint Case-B records (17.7%). Strong support thus did not override structural non-identifiability. The $H_1$ arm contributed 696 joint cases among 3,000 opportunities (23.2%), and the $H_2$ arm contributed 635 among 4,500 (14.1%); all 3,000 and 4,500 respective opportunities remained structurally rejected. These arm-specific rates, like the pooled rate, describe the prospectively enriched construction. The percentage characterizes the enriched design, not natural Case-B prevalence. The two populations are complementary and are not pooled: the general population identifies UPR, source risk, and Case A, whereas the targeted population isolates Case B.

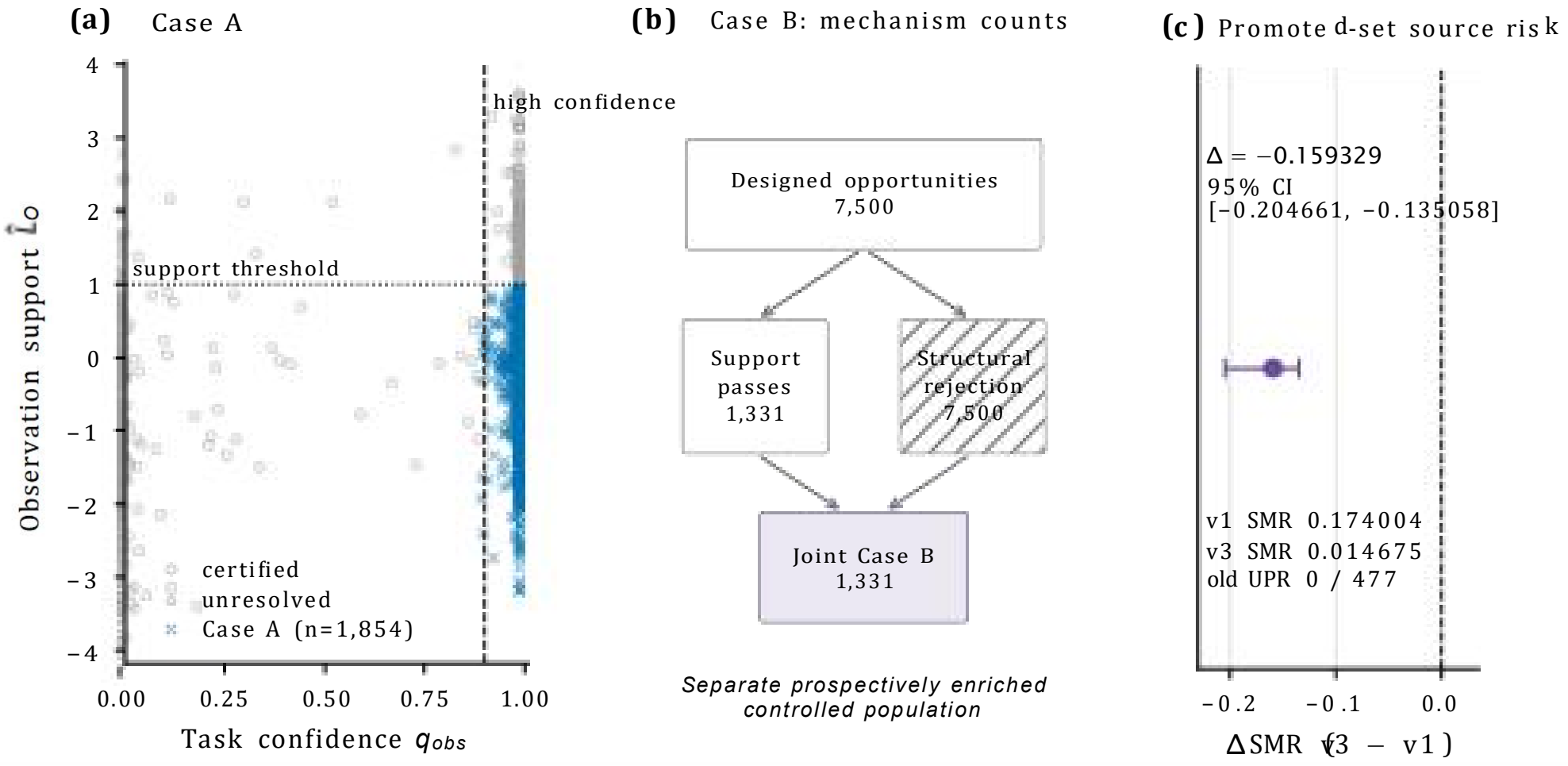

**Fig. 8 Confidence, observation support, structural certification, and source attribution are non-substitutable.** In the general held-out population, high task confidence co-occurs with insufficient observation support; marker form distinguishes certification state (a). Case B was tested in a separate prospectively enriched controlled population: all 7,500 opportunities were structurally rejected and 1,331 nevertheless passed observation support, demonstrating the mechanism rather than estimating natural prevalence (b). On the independently fixed promoted set, v3 sharply reduced old SMR while the paired 95% interval remained below zero and old UPR remained 0/477 (c).

## 7.4 Auxiliary evidence changes belief without changing observation warrant

Auxiliary interventions isolate belief updating from the observation basis of certification. Experiment 4 holds the observation realization and claim fixed while changing only the auxiliary evidence. Observation support was exactly invariant and structural certification had zero mismatches across neutral, aligned, and misleading arms (H4a). Nevertheless, aligned evidence raised fused belief relative to neutral evidence by +0.536 (paired seed-cluster 95% CI [+0.523, +0.549]; H4b). Auxiliary evidence therefore changed belief while leaving observation-relative epistemic status unchanged.

The same intervention preserved semantic non-relabeling exactly. The source readout made this separation empirically recoverable: every prescribed attribution metric improved, including a reduction in SMR from 0.304 to 0.007 (Table 3). Belief updating, observation warrant, and provenance thus remained distinct rather than collapsing into a single confidence response.

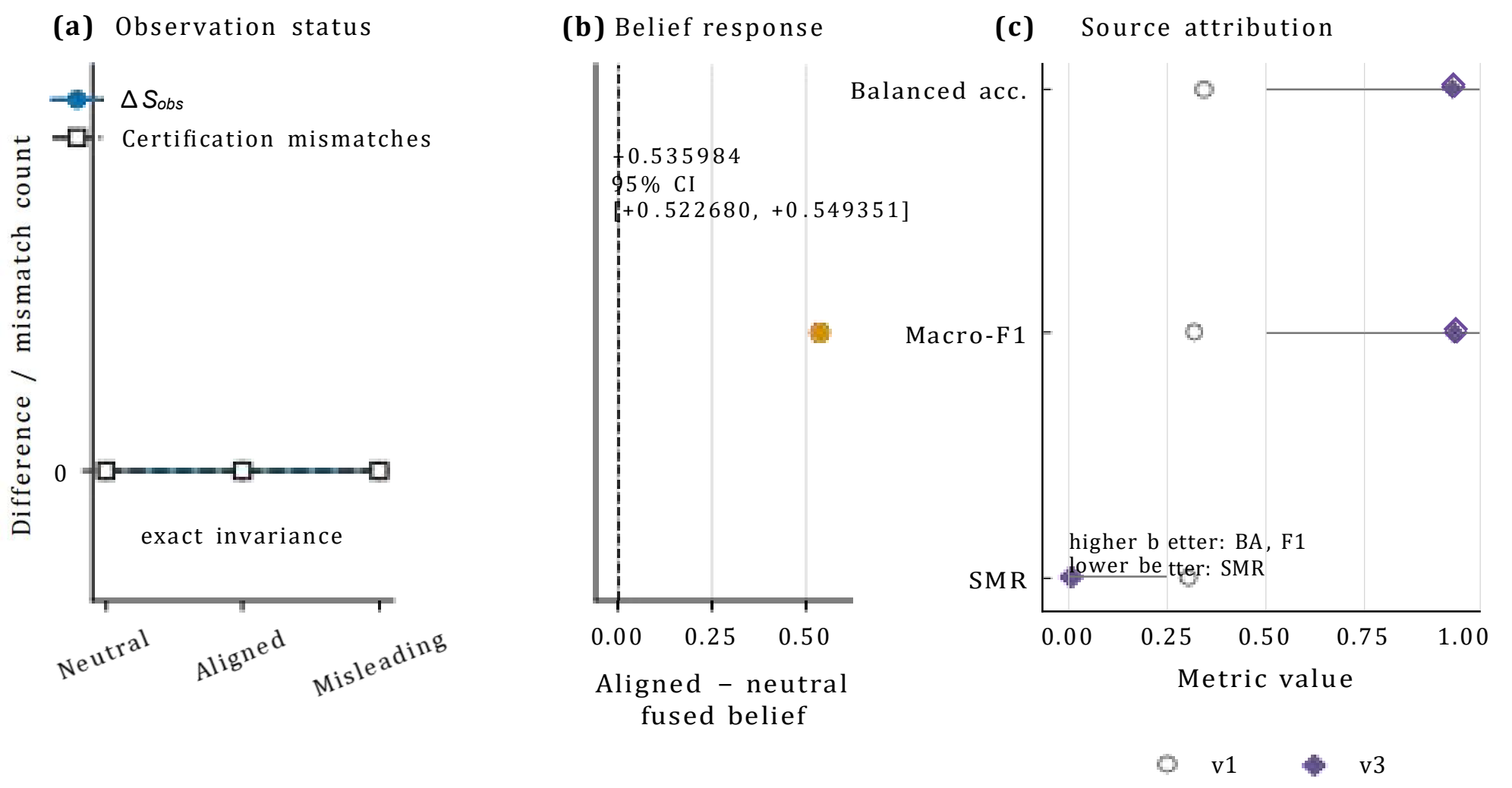

**Fig. 9 Auxiliary evidence changes belief and provenance without rewriting observation-relative status.** Auxiliary-only intervention leaves observation support and structural certification exactly invariant across neutral, aligned, and misleading arms (a). Aligned evidence raises fused belief relative to neutral evidence, with its paired 95% interval separated from zero (b). The v3 source readout improves balanced accuracy and macro-F1 while reducing SMR (c), making the same source separation recoverable from method-visible evidence.

## 7.5 Epistemic states propagate into knowledge dynamics

If certification and promotion depend on upstream epistemic state, acquisition loss should propagate into finite registries rather than remain a geometric diagnostic. Experiment 5 tests this consequence while keeping the claim family and routing rules fixed. The observation-certified registry contracted as 1381 → 613 → 271 → 1381, while the auxiliary-supported inference registry was nonempty, acquisition-responsive, and exactly restored as 1740 → 927 → 554 → 1740 (Figure 10). Restoration reproduced both baseline registries exactly.

The nonempty $I$ response was not driven by uncontrolled auxiliary-support false positives. At every acquisition condition there were 3,420 evaluator auxiliary-support opportunities and 3,423 method passes; precision was 0.999, recall 1.000, and false-positive rate 0.001; the full calibration diagnostics are reported below. These diagnostics rule out uncontrolled auxiliary-support false positives as the explanation for nonempty $I$. Degrading acquisition therefore propagated through certification

**Table 3** Source-related results for Experiments 3 and 4. Higher is better for accuracy, balanced accuracy, and macro-F1; lower is better for UPR and SMR.

| Endpoint | v1 | v3 | Change | 95% CI |
|---|---|---|---|---|
| Exp. 3 old UPR | — | 0/477 | — | — |
| Exp. 3 SMR | 0.174 | 0.015 | —0.159 | [—0.205, —0.135] |
| Exp. 3 accuracy† | 0.826 | 0.985 | — | — |
| Exp. 3 bal. accuracy† | 0.604 | 0.685 | — | — |
| Exp. 3 macro-F1† | 0.166 | 0.242 | — | — |
| Exp. 4 belief efect | reference | 0.536 | +0.536 | [+0.523, +0.549] |
| Exp. 4 accuracy† | 0.696 | 0.993 | — | — |
| Exp. 4 bal. accuracy† | 0.344 | 0.971 | — | — |
| Exp. 4 macro-F1† | 0.323 | 0.977 | — | — |
| Exp. 4 SMR† | 0.304 | 0.007 | —0.297 | — |

Old UPR and SMR report certification and source risk; accuracy, balanced accuracy, and macro-F1 report attribution; the Exp. 4 efect reports belief response. †Descriptive attribution summary; no confidence interval was reported.

and routing into both observation-certified knowledge and auxiliary-routed inference; restoring the operator recovered their baseline memberships. The v3 $I$ registry is a stricter operational realization of the broader inference concept in the original framework, not an assertion of set-definition identity. Exact restoration is a reproducibility result rather than a generalization claim.

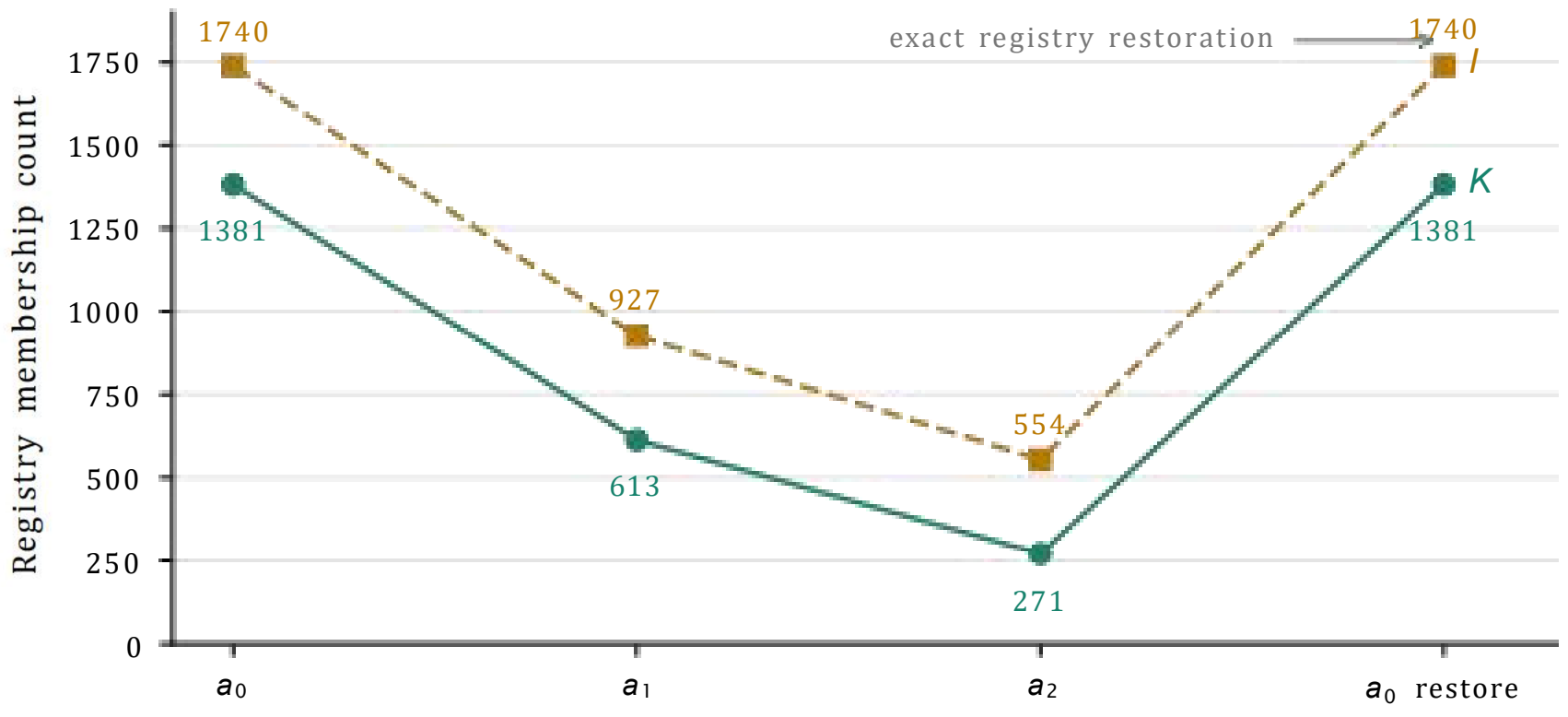

**Fig. 10 Acquisition degradation contracts both finite registries and restoration recovers them.** The $a_0 \to a_1 \to a_2 \to a_0^{\text{restore}}$ intervention removes observable distinctions and contracts both observation-certified $K$ and auxiliary-routed $I$; restoring the acquisition exactly reproduces both baseline registries. Here $I$ is the final v3 auxiliary-routing registry, whose membership rule is narrower than the broad operational inference registry in the original framework. Detailed transitions and auxiliary-support diagnostics appear in Figure 11 and Table 4.

Relative to baseline $H_0$, the $H_1$ transition contained 162 $K \to I$, 149 $I \to K$, 617 $I \to I$, and 310 new$\to I$ records. The $H_2$ transition contained 62 $K \to I$, 50 $I \to K$, 413 $I \to I$, and 141 new$\to I$ records. Restoration contained 1,740 $I \to I$ records and zero $K \to I$, $I \to K$, or new $\to I$ transitions. These are baseline-to-condition comparisons; they do not infer an unmeasured $H_1 \to H_2$ flow.

**Table 4** Finite-registry and auxiliary-support responses to the acquisition intervention. Opportunities and passes are evaluator-side opportunities and method-side auxiliary-support passes, respectively.

| Acquisition condition | $K$ | $I$ | Opportunities | Passes |
|---|---|---|---|---|
| $H_0$ | 1381 | 1740 | 3420 | 3423 |
| $H_1$ | 613 | 927 | 3420 | 3423 |
| $H_2$ | 271 | 554 | 3420 | 3423 |
| $H_0^{\mathrm{restore}}$ | 1381 | 1740 | 3420 | 3423 |

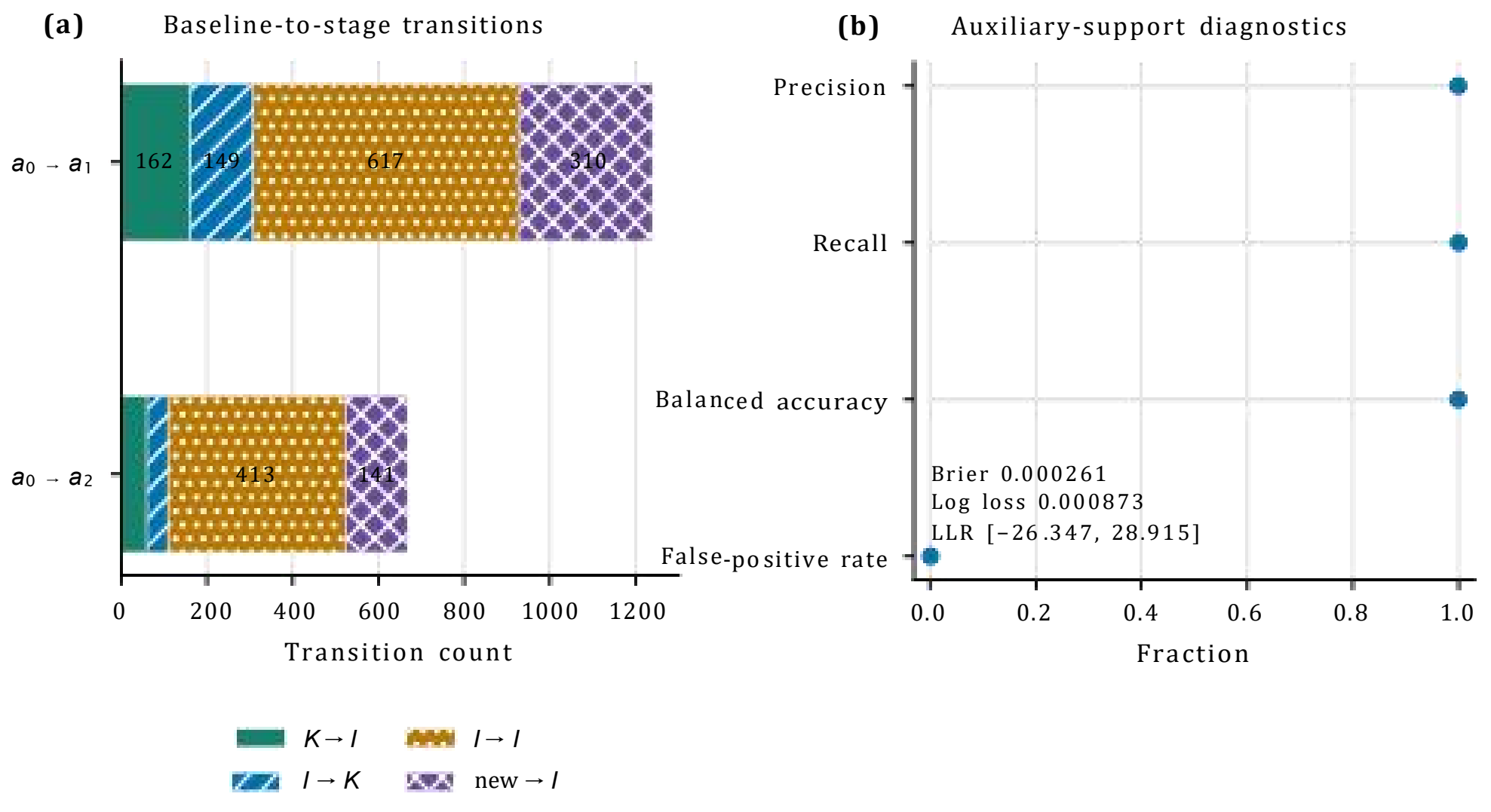

**Fig. 11 Registry transitions and auxiliary-support quality.** Baseline-to-degraded comparisons decompose membership changes without inferring an unmeasured $a_1 \to a_2$ flow (a). Auxiliary-support precision, recall, balanced accuracy, and false-positive rate show that the nonempty v3 $I$ registry is not produced by uncontrolled support positives (b). The diagnostic values are identical across the four acquisition conditions.

Across all four conditions, auxiliary-support precision was 0.999124, recall 1, balanced accuracy 0.999419, false-positive rate 0.001163, Brier score 0.000261, and log loss 0.000873; the auxiliary-evidence log-likelihood-ratio range was [__26.347186, 28.914878]. The identical diagnostic values across conditions reinforce that registry contraction arose from the acquisition-sensitive epistemic state rather than deterioration of the auxiliary-support readout.

## 7.6 Component interventions identify non-substitutable roles

Four interventions isolate different components and therefore do not form a common leaderboard (Table 5). Removing global analysis changed false certifications from 0/2 to 2/2 in the aliasing control. Removing attribution control increased SMR from 0.838000 to 0.891333. Removing support eligibility yielded the same RUPR, 0.550375, in both arms because the estimator was degenerate and passed every record; the null contrast does not show that support is unnecessary. Using the exact local bridge avoided 200,000 global checks with zero disagreements, a computational result restricted to the proved model class. For the associated historical headline comparator inference, the structural-promotion IPR comparison had raw $p = 9.999 \times 10^{-5}$ and Holm-adjusted $p = 2.9997 \times 10^{-4}$; the entangled-source comparison had raw and adjusted $p = 9.999 \times 10^{-5}$. These inferential results retain their original prespecified families and do not combine the role-specific interventions into a common score. At their separately matched $q = .20$ coverages, B5 reduced structural IPR from 0.250 for B3 to 0, and reduced entangled-source SMR from 0.654 for B-SRC to 0.357 (Figure 12). The two endpoints use different populations and external oracles and are therefore displayed as parallel contrasts rather than a common performance scale.

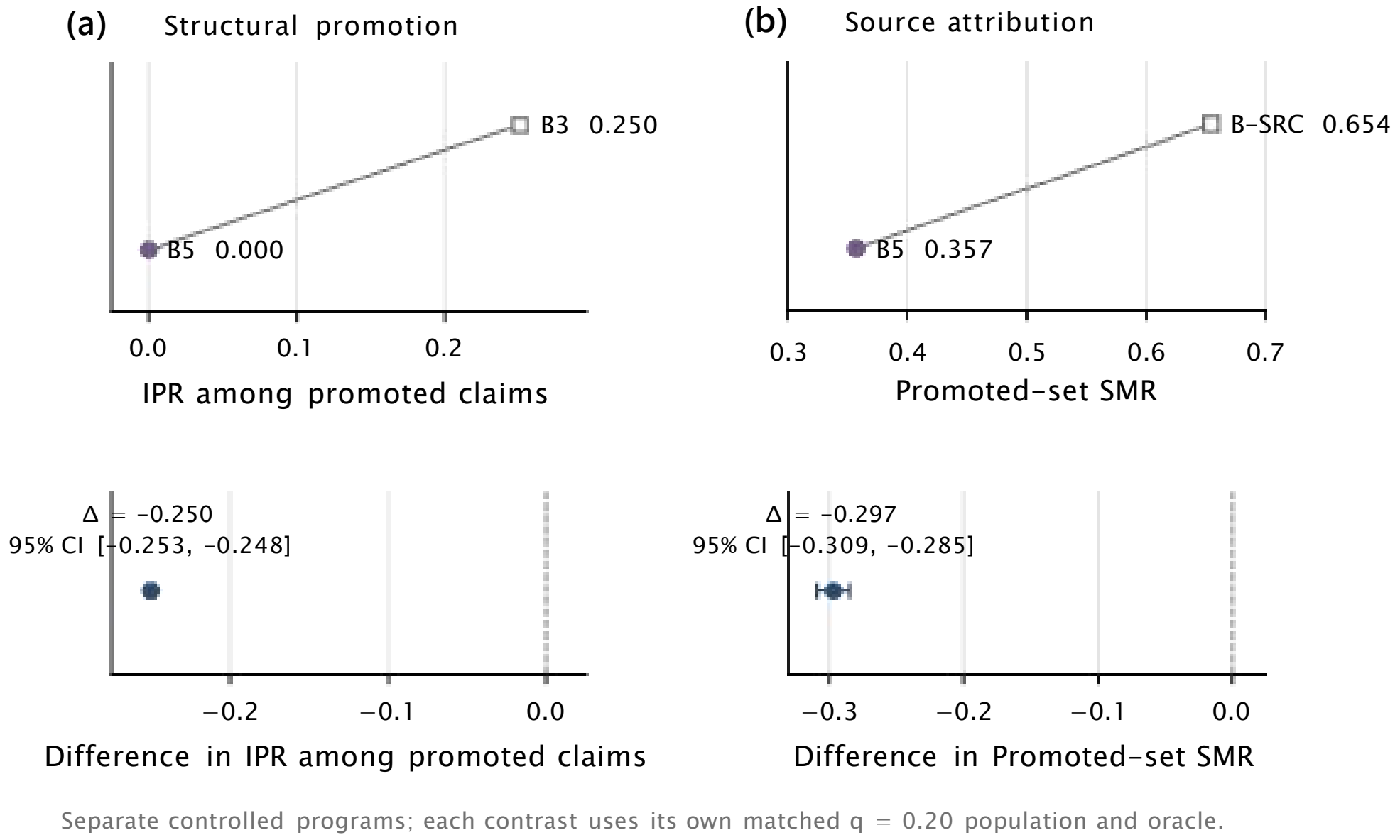

**Fig. 12 Matched-coverage comparisons isolate structural and source risk.** At separately fixed $q = .20$ coverage, the structural-promotion comparison measures IPR for B5 versus B3, whereas the entangled-source comparison measures promoted-set SMR for B5 versus the information-matched B-SRC comparator. Both paired intervals remain below zero, but the endpoints answer distinct scientific questions and use distinct evaluator oracles. The figure therefore supports role-specific controlled comparisons, not a common method leaderboard or a cross-program risk scale.

**Table 5** Component interventions isolate distinct scientific roles. Each interpretation is limited to its measured endpoint and controlled environment.

| Intervention | Scientific role | Observed efect | Interpretation | Boundary |
|---|---|---|---|---|
| Remove global analysis | Aliasing correctness | False certifications 0/2 → 2/2 | Global analysis is non-substitutable | Local Fisher is not intrinsically incorrect |
| Remove attribution control | Source semantics | SMR 0.838000 → 0.891333 | Source control reduces error here | Attribution is not universally solved |
| Remove support eligibility | Realized support | RUPR 0.550375 in both arms | All-pass support yields no contrast | Support is not unnecessary |
| Use exact local bridge | Qualified computation | 200,000 checks avoided; zero disagreements | Proved bridge saves checks | Not a nonlinear shortcut |

## 7.7 Typed integration exposes dependence and grouping boundaries

For a source-local coordinate map $x_m = \psi_m(\xi)$, GET transported Fisher geometry as $\Gamma_{m\to\xi} = J_m^T \Gamma_m J_m$ and reproduced the analytic geometry fixtures. Independent Gaussian source geometry added to joint Fisher 2. Under dependence $\rho = .5$, the marginal sum remained approximately 2 while exact joint Fisher was 4/3 (Figure 13); hence addition is assumption-dependent and says nothing about support-score fusion.

Referential grouping remained an empirical boundary. The initial connected-component rule achieved pairwise F1 0.952 and exact recovery 0.713 (95% CI [0.704, 0.722]). A subsequent prospectively specified one-to-one comparison achieved pairwise F1 0.982 and independent-qualification exact recovery 0.888 (95% CI [0.881, 0.894]). Its LCB 0.881 missed the declared 0.90 requirement, so the evidence did not establish grouping adequacy. One-to-one matching removed chain multiplicity but did not eliminate spurious associations. The same typed interface was exercised for v-GET, s-GET, and t-GET. This demonstrates module breadth, not referential efficacy, exact binding, grouping adequacy, or naturalistic end-to-end promotion.

**Table 6** Referential-grouping comparisons. The one-to-one rule improves pairwise and exact recovery but remains below the declared exact-recovery criterion.

| Association rule | Pairwise F1 | Exact recovery | 95% CI | Interpretation |
|---|---|---|---|---|
| Connected components | 0.951888 | 0.7134 | [0.7044,0.7221] | Chain associations reduce exact recovery |
| One-to-one partial matching | 0.982239 | 0.8875 | [0.8813,0.8936] | Improved, but LCB < 0.90 |

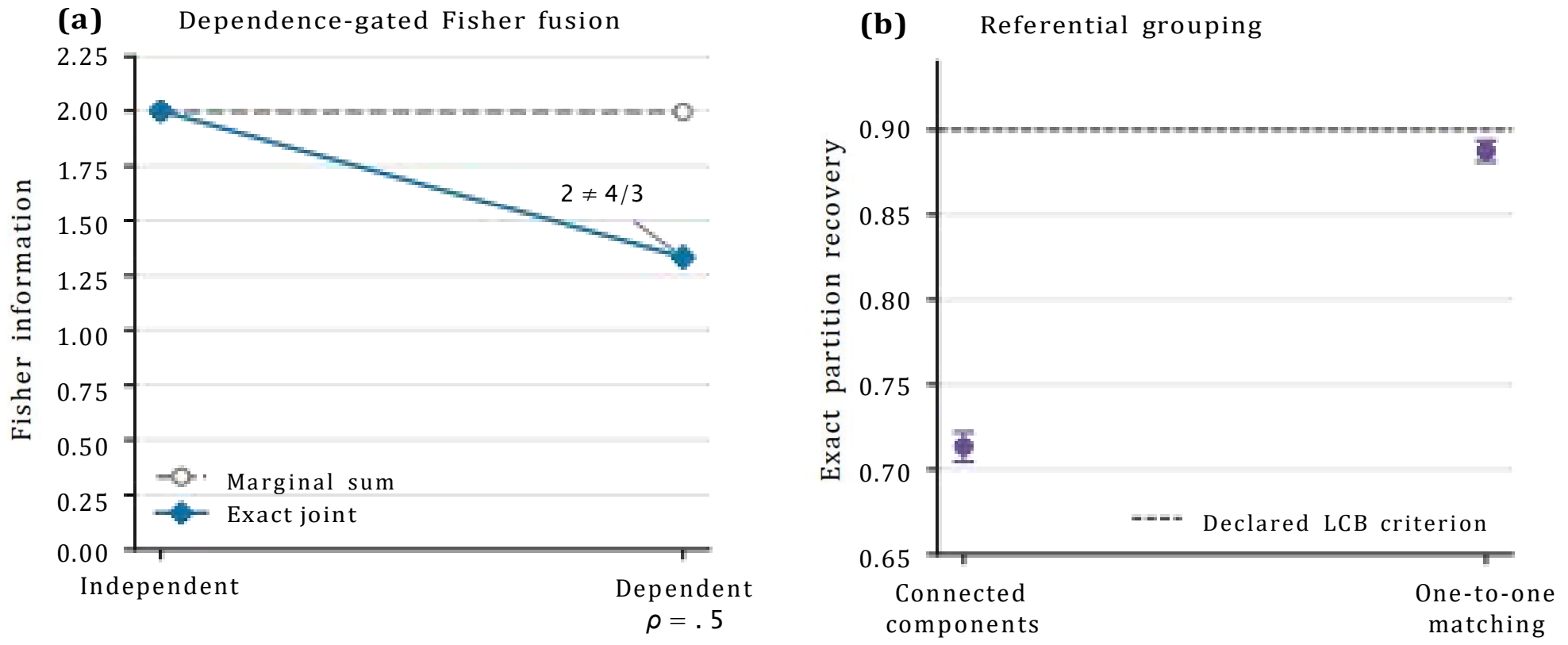

**Fig. 13 Typed integration depends on both evidence dependence and referential association.** Marginal Fisher addition agrees with the exact joint result under the independent law but not under the dependent law (a). One-to-one matching improves exact grouping over connected components, yet its interval remains below the declared lower-bound criterion (b). Geometry additivity therefore implies neither support additivity nor grouping adequacy.

## 7.8 High-fidelity simulation reveals compounding operational bottlenecks

The controlled experiments supply exact claim, distinction, and source semantics. To ask what changes when those premises must themselves be estimated, we applied the typed pipeline to ProcTHOR/AI2-THOR scenes with rendered multimodal observations. In the independent 40-house validation population, candidate coverage was 0.117, RH@32 was 0.107, and proposal-conditional grounding accuracy was 0.099. Referential grouping had zero exact recovery, no object-specific binding was verified, and the observation-support estimator rejected every candidate. Consequently final certification and $K$ were empty, while 33 records entered the auxiliary-routing registry.

These measurements locate a boundary rather than negate the controlled mechanism. As environmental realism increases, errors in grounding, grouping, binding, support, and attribution compound before promotion is reached. The simulation therefore functions as an operational stress test: it shows which controlled premises are difficult to recover from a multimodal perception pipeline, not that the underlying admissibility and source-separation principles fail.

## 7.9 Real-world data define evaluation boundaries

Real-world datasets reveal the limits of the available epistemic evaluation infrastructure. In the controlled program, structural and source relations are known by construction; in simulation, they must be estimated but remain evaluator-accessible. TbV and nuScenes ask whether plausible real-data proxies can stand in for missing claim-specific evidence semantics. Table 7locates these complementary tests within the evidence ladder, while Figure 14preserves their distinct quantitative questions without imposing a common performance scale. Neither study evaluates positive end-to-end natural-data efficacy.

### 7.9.1 TbV: computational decomposition is not epistemic provenance

The AV2 Trust, but Verify map-change study asked whether an exact computational decomposition of a multimodal classifier also provided a valid epistemic source relation. For the immutable late-fusion network, the agree-versus-disagree margin decomposed exactly as

$$\ell = \ell_{\text{sensor}} + \ell_{\text{map}} + b.$$

Opposite-signed branch contributions expose an architecture-native relation that the scalar margin cannot generally recover, making branch conflict a plausible prospective proxy rather than a validated epistemic fact. The controller made a positive map-agreement claim eligible only when the joint margin and both branch contributions had the positive direction; opposite branch signs were unresolved. This was a paired third-party-backbone case study, not an independent test of backbone generalization.

The decomposition was exact, but it did not provide the required epistemic proxy. Only 81/4459 records were promotion-eligible, so both $q = 0.10$ and $q = 0.20$ were unreachable. Branch conflict occurred on 90.2% of records: 91.3% when the released label was map agreement and 87.7% when it was disagreement. Treating conflict as a disagreement diagnostic gave balanced accuracy approximately 0.482 and $\phi \approx$ __0.057. Thus the branch relation did not track physical map disagreement in this interface.

This result neither excludes provenance in learned representations nor challenges the formal framework. It establishes a narrower boundary: without external semantic validation, exact architecture-native attribution is not epistemic provenance. More generally, computational saliency need not be faithful explanation, and feature relevance does not automatically acquire causal or epistemic meaning [40–42].

### 7.9.2 nuScenes: source availability is not claim-specific support

The nuScenes study strengthened source semantics by using physical acquisition metadata rather than neural internals. A direct-range eligibility contract used only camera/LiDAR acquisition mode, complete 10-sweep LiDAR history, timestamps, calibration, ego pose, and physical linkage. It did not use attention, hidden features, BEV embeddings, ground-truth visibility, ground-truth point counts, correctness labels, or neural attribution. The claim combined object existence, class, localization within 2 m of a same-class annotation, and matched-object distance below 30 m.

At primary coverage $q = 0.20$,

$$R_{\mathrm{B5}} = 0.982, \qquad R_{\mathrm{B\text{-}CONF}} = 0.971,$$

and

$$\Delta R = +0.010, \qquad 95\%\ \mathrm{CI} = [+0.004, +0.022],$$

using the preregistered paired 18-log bootstrap. The source-availability rule did not improve selective safety and was adverse relative to confidence-only selection in this comparison.

The absolute risk requires care. The prespecified universe comprised unthresholded post-decoder query–class detections, limited to the top 300, with predicted distance below 30 m. It was deliberately broader than a conventional high-confidence deployment population, and the endpoint combined detection, class, localization, and distance. Most errors reflected localization or the absence of a same-class object, rather than matched objects beyond the distance boundary. The result therefore does not show that LiDAR distance estimation failed or that LiDAR was harmful.

The combined universe contained 1,470,638 claims, of which 666,391 were B5-eligible. Matching global coverage required B5 to select 44.1% of its eligible population, reaching deeper into the camera–LiDAR score ranking than the confidence baseline. This rank-depth pattern describes the adverse comparison; it is not a post-hoc causal theorem.

The eligibility variable was constant across claims within an acquisition mode/keyframe. It established that a LiDAR window was available, not that raw returns geometrically bore on a particular promoted prediction. The boundary is precise: physical source availability did not imply claim-specific realized support in this operationalization. This is not a universal claim about the value of available modalities; multimodal reliability methods already recognize that availability, quality, and conflict require explicit treatment [10, 25].

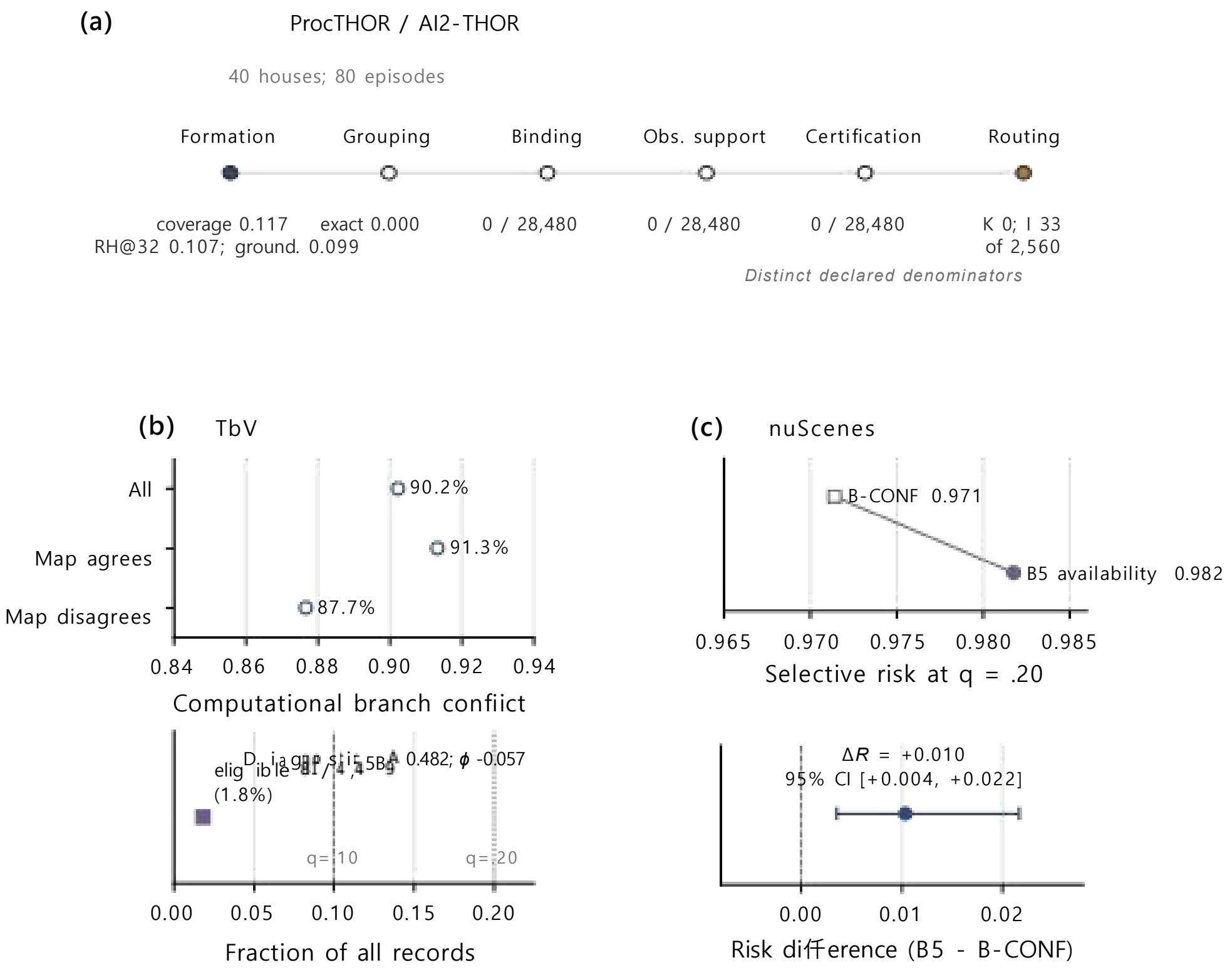

**Fig. 14 Operational realization and real-data proxies expose distinct epistemic boundaries.** In ProcTHOR/AI2-THOR, claim formation, grouping, binding, and observation support are measured under their declared denominators, showing how upstream estimation bottlenecks compound before certification and routing (a). In TbV, branch conflict is nearly ubiquitous, frozen coverage targets are unreachable, and the conflict diagnostic does not track released physical disagreement labels (b). In nuScenes, the physical availability selector has higher selective risk than confidence-only selection and the paired 95% interval for $\Delta R$ lies above zero (c). These panels represent simulated operational stress, a provenance-proxy boundary, and a realized-support boundary, respectively; they neither share a success scale nor establish positive real-world end-to-end efficacy.

Table 7 distinguishes positive controlled mechanism evidence from simulated operational stress and real-data boundary evidence; it does not place these settings on a common performance scale.

### 7.9.3 Combined boundary interpretation

The boundaries are complementary. TbV supplied claim-varying internal contributions but lacked an externally validated epistemic meaning. nuScenes supplied externally defined source identity and availability but not a claim-specific relation between realized measurements and each prediction. Within these operationalizations,

$$\text{source identity} \neq \text{source availability} \neq \text{claim-specific realized support.}$$

**Table 7** Evidence roles across controlled, simulated, and natural-data settings. Increasing realism changes which epistemic premises are known or must be estimated; TbV and nuScenes do not establish positive real-world end-to-end efficacy.

| Setting | Data type | Question tested | Evidence role | Primary conclusion | Scope boundary |
|---|---|---|---|---|---|
| Controlled v3 | Controlled generator | Do the declared mechanisms respond as predicted? | Controlled mechanism | Experiments 1–5 show the predicted responses | Controlled only |
| ProcTHOR / AI2-THOR | Simulated multi-modal scenes | What fails when premises must be estimated? | Simulated stress | Grounding, grouping, binding, and support bottlenecks compound | Not deployment efficacy |
| TbV | Real sensor/map data | Does branch decomposition supply epistemic provenance? | Real-data boundary | 81/4459 eligible; branch relation misses physical disagreement | Not positive selective efficacy |
| nuScenes | Real camera/LiDAR data | Does availability imply claim-specific support? | Real-data boundary | $\Delta R = +0.010$, 95% CI [+0.004, +0.022] | Not positive end-to-end efficacy |

Together they indicate that evidence semantics for natural-data certification should be claim-specific and externally grounded. They do not prove that any particular claim-specific operator will improve selective risk.

# 8 Discussion

## 8.1 Structured epistemic state is decision-relevant

The controlled results support a single mechanistic account. Acquisition determines which latent distinctions survive in the observation experiment, and admissibility depends on which of those distinctions a claim requires. Degradation therefore changes certification selectively rather than causing uniform confidence decay. The complementary Case-A and Case-B results expose the corresponding scalar insufficiency: confidence cannot supply missing observation support, and strong realized support cannot repair structural aliasing. A decision-relevant claim state must retain these distinctions rather than compress them into one score.

Source semantics add a further independent role. Aligned auxiliary information changed fused belief without changing observation support or structural certification, while explicit attribution reduced source error. An evidential contribution can therefore alter belief without acquiring the epistemic status of an observation. This is why source identity and lineage cannot be inferred from confidence alone.

The finite registries make the policy consequence of that structured state observable. As acquisition lost distinctions, both observation-certified $K$ and the narrower auxiliary-routing registry $I$ contracted; restoring the observation operator recovered their baseline memberships. The registries are thus downstream consequences of admissibility, support, and source-aware routing, not alternative measures of predictive accuracy.

Natural data expose the measurement boundary of this account. Controlled interventions can establish whether observation, auxiliary information, their combination, or neither supports a claim; ordinary task datasets rarely provide an equivalent oracle. TbV shows that exact model-internal decomposition need not carry externally valid epistemic source meaning. nuScenes shows that source availability need not establish that realized measurements support an individual claim. Source identity, acquisition availability, and claim-specific support are therefore distinct. The results support selected claim-level conformance relations, not a complete assurance ontology, a system-level competence model, or positive real-world end-to-end efficacy.

## 8.2 What increasing environmental realism changes

The controlled, simulated, and natural-data studies answer different questions. Controlled interventions identify mechanisms when claim and evidence semantics are known. The ProcTHOR/AI2-THOR stress test shows that estimating those premises introduces compounding grounding, association, binding, support, and attribution errors. TbV and nuScenes then show that real sensor datasets can expose useful boundary conditions without supplying all of the oracle relations needed for positive end-to-end epistemic evaluation. Controlled efficacy therefore does not by itself imply deployability.

Conventional task labels supervise whether a task claim $Y_j$ is correct. Source-aware epistemic evaluation ideally also observes a relation

$$R_{ij}^{GT} = \text{externally grounded relation between evidence } e_i \text{ and claim } c_j\,.$$

This is a benchmark requirement, not a theorem and not a claim that all existing datasets omit such information. Fact-verification and rationale benchmarks already annotate claim–evidence relations in other domains [31–33], and concurrent work studies claim-level specificity, certification, and multimodal attribution [34–36]. The narrower gap is that the two audited multimodal physical-sensor benchmarks did not natively provide the externally grounded claim–source relation required here. A useful schema could distinguish support, contradiction, irrelevance, and censored or unknown relations. It should be independent of the controller and should not turn absence of annotation into negative evidence.

## 8.3 Limitations

The guarantees are conditional on candidate inclusion and correct grounding. Proposal coverage and grounding must therefore be evaluated separately from certification conditional on inclusion; exact grounding remained weak in the controlled proposal study.

Reliable auxiliary-support and attribution estimates within the controlled generator do not imply equal reliability outside it. Although promoted-set SMR was low, Exp3 promoted-subset macro-F1 remained 0.242, leaving a selective-population limitation in absolute classwise attribution. The targeted Case-B rate is not a natural-prevalence estimate because the population was enriched prospectively for structurally unavailable claims. The v3 $I$ registry is also narrower than the original inference concept; its acquisition-responsive dynamics do not imply identical set definitions.

GET's generic typed interface and analytic geometry were valid, but neither grouping rule met the declared exact-recovery criterion. Exercising v-GET, s-GET, and t-GET demonstrated module breadth, not cross-modal referential grouping, exact binding, or positive naturalistic efficacy.

High-fidelity simulation localized compounding bottlenecks in grounding, grouping, binding, support, attribution, and promotion. TbV supplied computational rather than externally epistemic branch semantics, and nuScenes supplied source availability rather than claim-specific realized support. These boundaries do not support universal claims about learned provenance, physical sensors, or natural-data certification. Standard perception datasets also lack the complete structural, claim-support, and source-relation oracles required for the full evaluation used here. Positive real-world end-to-end efficacy is therefore not established.

Exact real-world observation fibres are generally unavailable, and symbolic certification remains restricted to declared finite or proved model classes. Fisher and CGM diagnostics depend on regularity, model, and coordinate choices; singular and globally aliased cases require global analysis. Finally, deterministic registry/geometry restoration is reproducibility under a restored operator, not generalization.

The epistemic-integrity positioning is also deliberately limited. This work does not estimate a latent epistemic competence scale, establish completeness or orthogonality of FASGL or validate it as a complete construct, fit a multidimensional item-response model, or test measurement invariance. It does not validate KnowBench or KnowProbe instruments, nor an ECO/EICI/EAL runtime-assurance stack. Those targets require separate instrument design and evidence beyond the criterion-referenced claim records studied here. The controlled results establish selected conformance relations under their declared generators; the natural-data studies remain proxy and oracle boundary evidence, and positive real-world end-to-end efficacy is not established.

## 8.4 Requirements for future claim–evidence benchmarks

Future evaluations should begin with the claim–evidence relation rather than infer it after observing task outcomes. Priorities include prospectively defined claim-specific evidence annotations, controlled source interventions, physically grounded claim–measurement relations, and explicit treatment of censored evidence relations. These requirements are more important than adding another confidence score or assuming that architectural branches inherit epistemic semantics.

# 9 Methods and Reproducibility

## 9.1 Controlled generators and claim families

The controlled program used prespecified finite claim libraries and external truths that were unavailable to the tested methods. Structural truth was computed from known observation fibres or proved model-class rules. Source truth was computed from prespecified observation and auxiliary neutralizations with conflict priority. The six controlled source classes were observation-only, auxiliary-only, redundant, joint-only, unsupported, and conflicting; these are properties of the declared generator and interventions, not an assertion that arbitrary learned systems expose the same labels.

The structural acquisition sequence was $H_0 \rightarrow H_1 \rightarrow H_2 \rightarrow H_0^{\mathrm{restore}}$. Structural interventions changed the row space of $H$, while the strength-only control held $H$ fixed and varied finite Gaussian noise. The separate degeneracy family was $\Gamma(\lambda) = \mathrm{diag}(1, .3, \lambda)$ over the prespecified grid $\lambda \in \{1, .3, .1, .03, .01, .003, .001, 0\}$.

## 9.2 Candidate proposal and grounding

The claim-formation study used six true referents per scene and a controlled five-template grammar. It performed exact candidate enumeration, applied the prespecified proposal score, selected deterministic Top-$K$ candidates, and evaluated referent recovery with an evaluation-only grounding oracle. Recall of hypotheses at $K$ measured whether a target survived proposal and selection; Grounding Accuracy measured exact referent recovery among proposed claims. Evaluation identities were not available to proposal scoring or grounding.

## 9.3 Structural oracles and admissibility procedures

The linear–Gaussian conditions used exact quotient, null-space, and row-space rules. The cubic-normal, cubic-Bernoulli, and two-sided $x^2$ cases supplied singular and global aliasing controls. $A_j^{GT}$ was computed externally and used only to score structural decisions. The internal procedure returned pass, fail, or unknown; unverified nonlinear cases were not converted into structural rejection. Descriptive rank, stabilized logdet, directional information, weak-subspace error, and CGM used the declared coordinate and scale conventions and did not enter the exact certificate outside a proved bridge.

## 9.4 Observation and auxiliary support estimation

The final controlled implementation, KNowLEDIzATIoN v3, used one canonical record schema across Experiments 3–5. Each method record stored immutable record, seed, scene, claim, candidate, and condition identities together with the historical seed-83201 task logit and

$$q_{\mathrm{obs}} = \mathrm{sigmoid}(\text{task–logit}).$$

Auxiliary support was estimated by a claim-conditioned two-layer ReLU network with widths [32, 32]. After calibration on the designated calibration split, its evidence score was $\widehat{L}_A = \mathrm{logit}(p_\theta(R_A = 1 \mid c, e_A)) - \mathrm{logit}(\pi_A)$, with the unchanged strict gate $\widehat{L}_A > 1$. Belief fusion used the fixed unit-evidence rule $\mathrm{logit}(q_{fused}) = \mathrm{logit}(q_{obs}) + \widehat{L}_A$.

The observation model, global/local admissibility, observation-support equation, and structural-certification rule were unchanged from Experiments 1–2.

## 9.5 Source-relation construction and attribution

The generator assigned the external six-state source relation through prespecified counterfactual observation and auxiliary neutralizations. Neither the state label nor its counterfactual bits were method inputs. Auxiliary neutralization was required to leave $S_{\mathrm{obs}}$ unchanged, and observation neutralization was required to leave $S_{\mathrm{aux}}$ unchanged. A symmetric squared-observation/sign-claim case served as a negative control for false observation support.

A separate unweighted two-layer ReLU readout with widths [32, 32] predicted the declared six-state source relation from method-visible source evidence. It was evaluated after legacy promotion membership was fixed and was not itself a promotion predicate. Canonical identities, confidence, support, certification, attribution, and registry routing were fixed before evaluator truth was joined.

## 9.6 Certification, promotion, and finite routing

The controlled B5 policy used structural pass and strict observation support $S_j^{\mathrm{obs}} > 1$ for certification eligibility. Source attribution was recorded and evaluated but was not an additional promotion predicate. Promoted records entered $K$; records satisfying the declared auxiliary-routing contract but not in $K$ entered the narrower v3 $I$ registry. The claim family and routing rules were fixed across the four acquisition conditions. Registry transitions were computed by canonical identity relative to baseline, not by matching numerical support tuples.

## 9.7 Predictive comparators and matched coverage

The primary promotion-safety comparison was B5 versus B3, and the primary entangled-source comparison was B5 versus B-SRC. B4 was a component control only where continuous geometry was defined; B6 was an oracle reference rather than a competitive method. The controlled promotion benchmark had no learned predictor. The separate learned-predictor comparison used raw confidence, temperature scaling, MC dropout, a five-member ensemble, and an exact posterior as predictive rather than structural or source comparators.

For target fraction $q$, each method selected exactly $k = \lfloor qN \rfloor$ eligible records when reachable under a common deterministic tie order; unreachable coverage was reported N/A. The two headline matched-coverage comparisons each used 20 complete held-out datasets with $N = 2400$, $q = .20$, and $k = 480$. Old UPR, IPR, RUPR, and SMR retained the distinct denominators and oracle dependencies defined in the technical appendix.

## 9.8 GET and grouping implementation

The generic evidence-token interface represented evidence, source identity, acquisition, grounding and referent links, coordinate systems, Fisher geometry, support, uncertainty, lineage, transformations, attribution, grouping, dependence/fusion state, and a derived ledger. Source-local geometry was transported by $\Gamma_{m\to\xi} = J_m^T \Gamma_m J_m$ . Independent and dependent analytic fixtures tested geometry composition. Referential groups were produced by connected components and by the prospectively specified one-to-one partial matching rule. v-GET, s-GET, and t-GET used the same typed adapter; the adapter did not itself compute certification or promotion.

## 9.9 ProcTHOR/AI2-THOR protocol

The high-fidelity simulation study applied the typed pipeline to rendered multimodal ProcTHOR/AI2-THOR scenes. Unlike the controlled generator, the scene study required claim proposal, grounding, referential grouping, object binding, and support to be estimated. The reported validation population comprised 40 independent houses. The study was an operational stress test and did not redefine the controlled structural or source oracles.

## 9.10 TbV and nuScenes protocols

The TbV study used the immutable AV2 Trust, but Verify late-fusion classifier. Its agree-versus-disagree margin was decomposed into sensor, map, and bias contributions. A positive map-agreement claim was eligible only when the joint margin and both branch contributions had positive direction; opposite branch signs were unresolved. The study used identical third-party backbone outputs for the compared rules.

The nuScenes study used immutable camera/LiDAR backbone outputs. Direct-range eligibility depended only on acquisition mode, complete 10-sweep LiDAR history, timestamps, calibration, ego pose, and physical linkage. It excluded attention, hidden features, BEV embeddings, ground-truth visibility or point counts, correctness labels, and neural attribution. The compound endpoint required object existence, class, localization within 2 m of a same-class annotation, and matched-object distance below 30 m. B5 and confidence selection used the same backbone outputs and prospectively fixed protocol.

## 9.11 Statistical analysis and reproducibility

Analytic oracle matrices and deterministic fibre enumerations were reported directly. Stochastic summaries used independent datasets as the inferential unit rather than claims within a dataset. The headline synthetic comparisons used 10,000-draw percentile bootstrap intervals and 10,000 paired sign-flip draws with the prespecified Holm correction. The final Exp3–5 comparisons used percentile bootstrap over held-out seed clusters. nuScenes used the preregistered paired 18-log bootstrap. Predictions and matched selections were fixed before evaluator-only truth was joined. Exact seeds, tie-order serialization, multiplicity contracts, and full-precision traceability are retained in the technical appendix.

# 10 Conclusion

KNowLEDIzATIoN separates questions that a prediction score alone cannot settle. Quotient theory identifies claims that are structurally constant on observation fibres; regularity-qualified Fisher geometry supplies a local diagnostic; and explicit support, lineage, attribution, certification, and registry interfaces prevent one status from standing in for another. Controlled interventions show that acquisition changes observable distinctions and claim-selective admissibility, that confidence and realized support cannot replace structural certification, and that auxiliary evidence can update belief without being relabeled as observation evidence. These distinctions propagate into finite $K$ and narrower operational $I$ registries that contract under acquisition degradation and recover under restoration. Together, the results establish the RQ1–RQ3 mechanisms at their declared controlled scope.

The progression toward realistic sensing also defines the boundary of that conclusion. High-fidelity simulation exposes grounding, grouping, binding, support, and attribution as compounding operational bottlenecks; TbV and nuScenes show that computational decomposition and source availability are incomplete proxies for claim-specific support. These studies provide simulated and real-data boundary evidence, not positive real-world end-to-end deployment efficacy.

The central distinction is

$$\begin{aligned} \text{being predicted} &\neq \text{being structurally identifiable} \\ &\neq \text{being supported by a source} \\ &\neq \text{being epistemically promotable.} \end{aligned}$$

The formalism and controlled evidence show that semantic separation enables errors in structural promotion, source attribution, and certification to be evaluated independently. The boundary results identify the additional semantics and supervision needed before stronger natural-data certification claims are warranted.

Within this bounded scope, KNowLEDIzATIoN provides a claim-level operational realization of epistemic-conformance principles; broader competence measurement and positive real-world assurance remain separate questions.

# A Formal Proofs and Assumptions

*Proof of Theorem 4.1* If the refinement holds, define $\tilde{c}([x]_\Omega) = c(x)$. Fibre constancy makes this well-defined. Conversely, factorization gives $q_\Omega(x) = q_\Omega(x') \Rightarrow c(x) = c(x')$. □

*Proof of Theorem 4.6* Positive-definite Gaussian noise gives equal laws exactly when $Hx = Hx'$. Thus fibre diferences are $N(H)$, and a linear claim is constant on those fibres exactly when $N(H) \subseteq N(C)$. Finally, $v^T \Gamma v = \|\Sigma^{-1/2} H v\|^2$, so $N(\Gamma) = N(H)$. □

The quotient theorem is set-theoretic and exact under the declared kernel. The local Fisher theorem additionally requires differentiability and the regular-fibre equality. The linear-Gaussian equivalence requires the stated linear domain and positive-definite covariance. Multimodal Fisher additivity requires score regularity and vanishing cross terms; it is not inferred from architectural modularity. Boolean composition preserves admissibility only for already fibre-constant primitives.

# B Detailed Implementation Specifications

The controlled ontology used five relation/attribute templates. Candidate enumeration, proposal scoring, deterministic Top-$K$, and grounding used no evaluation identity. The synthetic program used prespecified finite claim libraries and external truth unavailable to tested methods. Structural truth was computed from known fibres or exact class rules. Source truth was computed from prespecified observation/auxiliary neutralizations with conflict priority. Method inputs excluded both oracles, which were joined only after decisions and matched-coverage selections were fixed. Record ties were resolved by method-independent preprediction identifiers, and primary summaries were independently recomputed from raw records.

The six controlled source classes were observation-only, auxiliary-only, redundant, joint-only, unsupported, and conflicting. They are properties of the declared generator and interventions, not an assertion that arbitrary learned systems expose the same labels. The GET adapter preserved source presence, support, geometry, dependence/fusion, grouping, lineage, and attribution as separate fields and did not itself compute admissibility, certification, or promotion.

# C Exact Comparator and Endpoint Contracts

# D Exact Statistical and Reproducibility Contracts

Canonical record IDs were allocated before scientific assignment. The hash omitted method, oracle label, prediction, and file position; ties therefore used a common method-independent order. For exact reproduction, the historical UTF-8 ordering identifiers were Gate3FreezeV1.1/PROMO-1/seed/record_id and Gate5FreezeV1.0/ENT-SRC-1/seed/record_id. Target counts used $\lfloor \text{Decimal}(q)N \rfloor$. The structural-promotion study's Holm correction covered its preregistered non-oracle comparator families; its primary IPR raw and adjusted $p$ values were $9.999 \times 10^{-5}$ and $2.9997 \times 10^{-4}$. The entangled-source study's single primary comparison had raw and adjusted $p = 9.999 \times 10^{-5}$.

# E Full-Precision Traceability

For Experiment 3, the H0-neutral population contained 6,000 records, the common confidence universe contained 1,200, and 477 were actually promoted. Historical UPR was 0. The paired old-SMR difference was __0.159329 with 95% CI [__0.204661, __0.135058]. The independent Case-B addendum contained 7,500 prospectively enriched opportunities: all were rejected, 1,331 passed observation support, and hence 1,331 instantiated Case B. This 17.7467% rate describes the enriched construction, not natural prevalence. Exact final source comparisons are consolidated in main-text Table 3.

**Table 8** Claim-level epistemic-control interface. "Internal" denotes a method-visible or method-computed quantity; evaluator-only truths are unavailable to the method.

| Object | Symbol / type | Meaning | Method visibility |
|---|---|---|---|
| Claim | $c_j \in L$ | Declared proposition or function | Declared input |
| Exact admissibility | $A_j(\Omega) \in \{0, 1\}$ | Fibre constancy $Q_\Omega \preceq Q_{c_j}$ | Structural property |
| Structural truth | $A_j^{GT} \in \{0, 1\}$ | Exact generator/model-class value | Evaluator only |
| Structural procedure | $D_j^A \in \{\text{pass, fail, unknown}\}$ | Sound internal result about admissibility | Internal output |
| Observation support | $S_j^{\text{obs}}$ | Claim-relative evidence from Ω | Internal / declared |
| Auxiliary support | $S_j^{\text{aux}}$ | Claim-relative evidence from an auxiliary source | Internal / declared |
| Realized-support truth | $R_j^{GT} \in \{0, 1\}$ | External record-level support relation | Evaluator only |
| Estimated support | $\hat{R}_j$ or $\hat{L}_{O,j}$ | Method-side estimate | Internal output |
| Lineage | $\Lambda_j \in \mathsf{L}$ | Source identifiers, transforms, and computation path | Internal record |
| Source attribution | $\hat{G}_j$ | Source relation credited with warrant | Internal output |
| Source truth | $G_j^{GT} \in G$ | External counterfactual source class | Evaluator only |
| Certification | $\hat{C}_j$ | Three-state structural/support decision | Internal output |
| Promotion | $P_j \in \{0, 1\}$ | Operational release under the declared policy | Internal output |

For Experiment 4, neutral, aligned, and misleading arms held world/task state, observation realization, acquisition, structure, and claim fixed. Observation support was exactly invariant and structural certification had zero paired mismatches. The aligned-minus-neutral fused-belief effect was +0.535984 with paired 95% CI [+0.522680, +0.549351]. Semantic non-relabeling was exact. The v1-to-v3 attribution comparison was accuracy 0.695667 → 0.992944, balanced accuracy 0.344482 → 0.971140, macro-F1 0.322816 → 0.977406, and SMR 0.304333 → 0.007056.

For Experiment 5, auxiliary-support precision was 0.999124, recall 1, balanced accuracy 0.999419, false-positive rate 0.001163, Brier score 0.000261, and log loss 0.000873 at each condition. The frozen auxiliary-evidence range was [__26.347186, 28.914878]. These values duplicate the Results traceability and introduce no appendix-only conclusion.

**Table 9** Synthetic comparator contracts. All matched selections share a record universe and deterministic tie order; evaluator truth is not an input.

| Method | Inputs / source access | Ranking and eligibility | Excluded information |
|---|---|---|---|
| B1 | Observation likelihood, auxiliary signal, prescribed posterior | Posterior probability; threshold or top-$k$ | Structural certificate; source output |
| B2 | Declared-reference $S_{\text{obs}}$ | $S_{\text{obs}} > 1$; rank by $S_{\text{obs}}$ | Auxiliary/source states; global certificate |
| B3 | Observation likelihood, auxiliary signal, prescribed posterior | Posterior risk/entropy; exact top-$k$ | Structural certificate; source output |
| B4 | Support and continuous local geometry where defined | Local pass and $S_{\text{obs}} > 1$ | Global/model-class certificate |
| B-SRC | Factual $(L_O, L_A)$ and posterior | Strongest signed component; exact top-$k$ | Six-state source relation |
| B5 | Factual evidence and allowed structural procedure | Structural pass and $S_{\text{obs}} > 1$ | External $A_j^{GT}$, $G_j^{GT}$, latent class |
| v3 | Canonical confidence, observation/auxiliary/source evidence | Legacy promotion; attribution evaluated afterward | Evaluator truth; changes to Exp. 1–2 machinery |

**Table 10** Endpoint semantics. Historical old UPR is an internal-certification quantity; IPR and RUPR use distinct external oracles and are not substitutes for it.

| Endpoint | Denominator / reference | Scientific meaning |
|---|---|---|
| Old UPR | Actually promoted claims; internal certification | Promoted fraction whose internal state is not certified |
| IPR | Promoted claims; $A_j^{GT}$ | Structural inadmissibility risk |
| RUPR | Promoted claims; $R_j^{GT}$ | Realized-support risk |
| SMR | Promoted claims; $G_j^{GT}$ | Strict source-attribution error |
| $\lvert A_H \rvert$/CSET | Exact finite-library oracle and state | Structural contraction and restoration |
| $K_t/I_t$ | Finite controller records | Promoted and auxiliary-inference membership |
| Fisher/TV | Analytic fixed-map quantities | Operational-strength diagnostics, not certificates |

# Data Availability

*Author-completion placeholder: identify the repositories, accession information, licensing conditions, and access restrictions for every dataset generated or analysed in this study.*

**Table 11** Preregistered statistical and selection contracts for the headline synthetic comparisons.

| Study | Held-out/analysis seeds | Coverage | Total ordering | Inference |
|---|---|---|---|---|
| Structural-promotion safety | 23001–23020; 29901 | $N$ = 2400 per seed; $q$ = .20; $k$ = 480 | $(-s, t_j, \text{id})$; common SHA-256 order | Complete-seed paired unit; 10,000 bootstrap and sign-flip draws; percentile 95% CI; preregistered Holm family |
| Entangled-source attribution | 43001–43020; 49901 | $N$ = 2400 per seed; $q$ = .20; $k$ = 480 | $(-s, t_j, \text{id})$; common SHA-256 order | Complete-seed paired unit; 10,000 bootstrap and sign-flip draws; percentile 95% CI; Holm family size one |
| Final v3 Exp3–5 comparison | 93401–93420 | H0-neutral $N$ = 6000; $q$ = .20 universe = 1200; actually promoted = 477 | Descending $q_{obs}$; SHA-256 tie-break | 10,000-draw percentile bootstrap over held-out seed clusters; predictions fixed before oracle join |
| Targeted Exp3 Case B | 93501–93520; 25 scenes each | 7,500 prospectively enriched $H_1/H_2$ opportunities | Design-side row-space membership before prediction | Fixed nonzero-mechanism criterion; descriptive enriched rate, not prevalence inference |

# Code Availability

*Author-completion placeholder: identify the repository, persistent identifier, software version, licence, and any access restrictions for the code underlying the reported results.*

# Author Contributions

*Author-completion placeholder: provide contribution statements for P.-H.H., L.P., Y.M., and Y.J. using author initials and confirm that all authors approved the manuscript.*

# Competing Interests

*Author-completion placeholder: provide the authors' financial and non-financial competing-interests declaration.*

## Funding

*Author-completion placeholder: identify all funders and grant numbers and state the funders' roles, or state that the study received no funding.*

## Acknowledgements

*Author-completion placeholder: add any acknowledgements that satisfy the journal's requirements, or remove this section if none apply.*